\documentclass[twocolumn]{aastex6}
\usepackage{amssymb}
\usepackage{graphicx}

\begin{document}

\title{How Density Environment Changes the Influence of the Dark Matter-Baryon
Streaming Velocity on the Cosmological Structure Formation}

\author{Kyungjin Ahn\altaffilmark{1}}
\affil{Department of Earth Sciences, Chosun University, Gwangju 61452,
  Korea}
\altaffiltext{1}{kjahn@chosun.ac.kr}

\begin{abstract}
We study the dynamical effect of relative velocities between dark
matter and baryonic fluids, which remained supersonic after the epoch
of recombination. The impact of this supersonic motion on the formation
of cosmological structures was first formulated by \citet{Tseliakhovich2010},
in terms of the linear theory of small-scale fluctuations coupled
to large-scale, relative velocities in mean-density regions. In their
formalism, they limited the large-scale density environment to be those
of the global mean density. We improve on their formulation by allowing
variation in the density environment as well as the relative velocities.
This leads to a new type of coupling between large-scale and
small-scale modes. 
We find that the small-scale fluctuation grows in a biased way: faster
in the overdense environment and slower in the underdense environment.
We also find that the net effect on the global power spectrum of the
density fluctuation is to boost its overall amplitude from the prediction
by \citet{Tseliakhovich2010}. Correspondingly, the conditional mass
function of cosmological halos and the halo bias parameter are both
affected in a similar way. The discrepancy between our prediction
and that by \citet{Tseliakhovich2010} is significant, and therefore
the related cosmology and high-redshift astrophysics should be revisited.
The mathematical formalism of this study can be used for generating
cosmological initial conditions of small-scale perturbations in generic,
overdense (underdense) background patches.
\end{abstract}

\keywords{cosmology: theory --- dark ages, reionization, first stars
--- surveys}

\section{Introduction}
\label{sec:Introduction}

The $\Lambda$-cold dark matter ($\Lambda$CDM) scenario, combined
with the theory of cosmic inflation, is the successful, concurrent
model describing the past and the present of our universe, consistent
with a wide range of observations. In this scenario, cosmological
structures grow out of an extremely uniform density field but with
tiny fluctuations that are seeded by the cosmic inflation. The growth
of the CDM density fluctuations and the growth of baryon density fluctuations
are not in perfect synchronization, because baryons were tightly coupled
to photons before the epoch of recombination and thus their motion
was different from the motion of CDM which only reacts to gravity.
Only after recombination baryons gradually decoupled from photons, and 
followed the motion of the CDM under gravity.

Cosmological observations have verified the $\Lambda$CDM scenario
in scales large enough to make the baryonic physics almost irrelevant
(e.g. \citealt{2011ApJS..192...18K}; \citealt{Reichardt2012} \citealt{2015arXiv150201589P}).
However, once in the regime where the baryonic physics becomes important,
the growth of baryon fluctuations is affected by hydrodynamics and
the growth of the CDM is affected by the gravitational feedback from
the baryon fluctuations. Some of the usual assumptions that are made
for treating very large scales, therefore, should be taken carefully
or modified when treating relatively small scales. For example, \citet{Naoz2005}
improved on the previous estimation of the linear density power spectrum
in small scales, by replacing the usual assumption made in cosmology
that the sound speed of baryons is uniform in space with the fact
that the sound speed fluctuates in space in small scales. They showed
that more than $\sim10\,\%$ change occurs in the baryon density power
spectrum and even more change in the baryon temperature power spectrum. 
A sheer inclusion of the sub-dominant, yet
non-negligible baryonic component in the analysis changes the
prediction on
the matter density power spectrum at a
few percent level even in large scales, as was shown in the
framework of the high-order perturbation theory
\citep{Shoji2009,Somogyi2010}.

Similarly, the relative velocity (``streaming velocity'') between baryons
and the CDM after the recombination should also be considered carefully
in cosmology. 
\citet[TH hereafter]{Tseliakhovich2010}, for the first time, properly
calculated the growth of  small-scale density fluctuations under
the influence of the streaming velocity. Small-scale fluctuations are
coupled to large-scale streaming velocity fields which are coherent over
a few comoving Mpc scale. This has its own spatial fluctuation with $\sim30\,$km/s
standard deviation at the epoch of recombination, and then decays
in proportion to the inverse of the scale factor $a$. Its impact
on the small-scale structure formation is non-negligible because the
streaming velocity remains supersonic (until the intergalactic medium
is strongly heated). This then leads to the suppression of small-scale
matter density fluctuations. The wave-modes that are the most strongly
affected are around ${\bf k}\sim200/$Mpc, and the impact is on the
matter power spectrum, the conditional mass function and the halo
bias parameter, to name a few (TH).

Subsequent studies have considered the impact of the streaming velocity
in the perspective of both cosmology and astrophysics. The boost of
amplitude and the shift of the peak of the baryonic acoustic oscillation
(BAO) feature due to the streaming velocity, in the gas intensity
mapping or the galaxy survey, were intensively investigated (\citealt{Dalal2010};
\citealt{Yoo2011}; \citealt{McQuinn2012}; \citealt{Slepian2015};
\citealt{Lewandowski2015}; \citealt{Blazek2015}; \citealt{Schmidt2016}).
They find that both high-redshift (e.g. \citealt{Dalal2010}; \citealt{McQuinn2012})
and low-redshift (e.g. \citealt{Yoo2011}) surveys will be affected.
The impact of the streaming velocity on BAO may be separated out from
the impact of the matter density itself (\citealt{Slepian2015}),
which is important because otherwise it will become another nuisance
parameter in cosmology. The astrophysical impact of the streaming
velocity has been investigated with focus on the formation of the
nonlinear structure such as cosmological halos and stellar objects
(\citealt{Maio2011}; \citealt{Stacy2011}; \citealt{Greif2011};
\citealt{Tseliakhovich2011}; \citealt{Naoz2012}; \citealt{Fialkov2012};
\citealt{O'Leary2012}; \citealt{Bovy2013}; \citealt{Richardson2013};
\citealt{Tanaka2013}; \citealt{Naoz2014}; \citealt{Popa2015}; \citealt{Asaba2016}).
Most studies indicate that the formation of minihalos (roughly in
the mass range $M=[10^{4}-10^{8}]\, M_{\odot}$) and the formation
of stellar objects in them are suppressed. It may induce baryon-dominated
objects such as globular clusters (\citealt{Naoz2014}; \citealt{Popa2015}),
but the actual star formation process leading to globular clusters
has yet to be simulated. It may be responsible even for the generation
of the primordial magnetic field (\citealt{Naoz2013}). Because minihalos
are the most strongly affected among cosmological halos and they are
responsible for the early phase of the cosmic reionization process,
how they change the high-redshift 21-cm background is also of a prime
interest (\citealt{Visbal2012}; \citealt{McQuinn2012}; \citealt{Fialkov2013}). It is noteworthy that some of these numerical simulation results (\citealt{Maio2011};
\citealt{Stacy2011}; \citealt{Greif2011}), which are based on the
initial condition generated by the usual Boltzmann solver such as
the Code for Anisotropies in the Microwave Background (CAMB: \citealt{Lewis2000}),
need to be re-examined. This is because the impact of the streaming
velocity is cumulative and inherent even at $z\sim200$ (\citealt{McQuinn2012}), at which
or later these simulations start with a streaming velocity implemented by hand.

The original formalism by TH has been re-investigated in terms of
the high-order perturbation theory in the wave-number space (``${\bf k}$-space''
henceforth), and some ``missing terms'' previously neglected were
found important (\citealt{Blazek2015}; \citealt{Schmidt2016}). Basically,
for the large-scale modes responsible for the streaming velocity (${\bf k}\sim[0.01-1]/{\rm Mpc}$),
TH used a trivial solution for the evolution of the streaming velocity
($V_{{\rm bc}}$) and the density ($\Delta_{{\rm c}}$ and $\Delta_{{\rm b}}$
being overdensities of the CDM and baryons in large scale, respectively)
environment: $V_{{\rm bc}}\propto a^{-1}$, $\Delta_{{\rm c}}=\Delta_{{\rm b}}=0$.
Treating this as a new 0th-order solution to the perturbation equations,
they then examined how the perturbation in small scales grows. In
doing so, they treated $V_{{\rm bc}}$ as a spatial quantity and perturbation
variables in small scales as ${\bf k}$-space quantities. This
is basically a high-order perturbation theory, coupling large-scale
mode ($V_{{\rm bc}}$) and the small-scale modes ($\delta_{{\rm c}}$
and $\delta_{{\rm b}}$, small scale CDM and baryonic overdensities,
respectively). However, $V_{{\rm bc}}$ is tightly linked to the fluctuating
$\Delta_{{\rm c}}$ and $\Delta_{{\rm b}}$ through the density continuity
equation, and thus the trivial solution adopted by TH cannot be used
for generically 
overdense and underdense regions. The continuity equation connects
the divergence of $V_{{\rm bc}}$ to $\Delta_{{\rm c}}$ and
$\Delta_{{\rm b}}$, and \citet{Schmidt2016} finds 
that the divergence of $V_{{\rm bc}}$ is indeed an important term that
one should not ignore. 
\citet{Blazek2015} also works on the generic basis of non-zero
$\Delta_{{\rm c}}$ and $\Delta_{{\rm b}}$.

We improve on the formalism of TH by also considering the non-zero
overdensities. Toward this end, different from \citet{Blazek2015}
and \citet{Schmidt2016}, we inherit the original method by TH 
and focus on the impact on small-scale modes:
large-scale $V_{{\rm bc}}$ is treated as the spatial quantity
and small-scale overdensities 
$\delta_{{\rm c}}$ and $\delta_{{\rm b}}$ are treated as the ${\bf k}$-space
quantity in the perturbation analysis. Most importantly, we explicitly
include generically ``non-zero'' $\Delta_{{\rm c}}$ and $\Delta_{{\rm b}}$
as a new set of spatial quantities, and consequently the divergence
of CDM and baryon velocities as well. We find that this leads to a
set of mode-mode coupling terms, including the velocity divergence-density
coupling. We carefully include all the coupling terms to the leading
order in our perturbation analysis. We also include the baryonic physics,
namely fluctuations in the sound speed (\citealt{Naoz2005}), the
gas temperature and the photon temperature. Our formalism is also
suitable for generating initial conditions for N-body+hydro numerical
simulations. Because the new set of mode-mode couplings is imprinted
in the initial condition, the initial condition generator considering
the streaming-velocity effect by \citet{O'Leary2012}, CICsASS,
should also be improved on if one were to numerically simulate the
structure formation inside overdense or underdense regions.

This paper is organized as follows. After the introduction, we lay
out the basic formalism and describe the statistics of large-scale
fluctuations in Section \ref{sec:formalism}. In Section \ref{sec:Result},
we show results on the matter density power spectrum, the conditional
halo abundance and the halo bias as applications of the formalism.
We conclude this work in Section \ref{sec:Discussion} with a summary,
discussion and future prospects.
Some details left out in the main body are described in Appendices.

\section{Formalism and Numerical Method}
\label{sec:formalism}

\subsection{Fluctuation under non-zero overdensity and relative velocity: perturbation
formalism}
\label{sub:perturbation}

We start from a set of equations for perturbations of relevant physical
variables. All the $k$ modes of interest are in sub-horizon scale such that
the Newtonian perturbation theory holds. Let us define the overdensity
$\delta_{j}\equiv\left(\rho_{j}-\bar{\rho}_{j}\right)/\bar{\rho}_{j}$,
where the subscript $j$=\{c,~b\} denotes either the CDM (c) or the
baryonic (b) component, and $\theta_{j}\equiv\left(1/a\right)\nabla\cdot{\bf v}_{j}$
where $a$ is the scale factor and ${\bf v}_{j}$ is the proper peculiar
velocity of component $j$. When the universe is in the regime where
we can ignore fluctuations of photons and neutrinos due to their rapid
diffusion after recombination, we have (e.g. \citealt{Bernardeau2002};
TH)

\begin{eqnarray}
\frac{\partial\delta_{{\rm c}}}{\partial t} & = & -a^{-1}{\bf v}_{{\rm c}}\cdot\nabla\delta_{{\rm c}}-a^{-1}(1+\delta_{{\rm c}})\nabla\cdot{\bf v}_{{\rm c}},\nonumber \\
\frac{\partial{\bf v}_{{\rm c}}}{\partial t} & = & -a^{-1}\left({\bf v}_{{\rm c}}\cdot\nabla\right){\bf v}_{{\rm c}}-a^{-1}\nabla\phi-H{\bf v}_{{\rm c}},\nonumber \\
\frac{\partial\delta_{{\rm b}}}{\partial t} & = & -a^{-1}{\bf v}_{{\rm b}}\cdot\nabla\delta_{{\rm b}}-a^{-1}(1+\delta_{{\rm b}})\nabla\cdot{\bf v}_{{\rm b}},\nonumber \\
\frac{\partial{\bf v}_{{\rm b}}}{\partial t} & = & -a^{-1}\left({\bf v}_{{\rm b}}\cdot\nabla\right){\bf v}_{{\rm b}}-a^{-1}\nabla\phi-H{\bf v}_{{\rm b}}-a^{-1}c_{s}^{2}\nabla\delta_{{\rm b}},\nonumber \\
\nabla^{2}\phi & = & 4\pi Ga^{2}\bar{\rho}_{m}\delta_{m},\label{eq:master}
\end{eqnarray}
where $a$ is the scale factor, $\nabla$ is the gradient in the comoving
frame, $\bar{\rho}_{m}=f_{{\rm c}}\bar{\rho}_{{\rm c}}+f_{{\rm b}}\bar{\rho}_{{\rm b}}$,
$\delta_{m}=f_{{\rm c}}\delta_{{\rm c}}+f_{{\rm b}}\delta_{{\rm b}}$,
$f_{{\rm c}}=\Omega_{{\rm c},0}/(\Omega_{{\rm c},0}+\Omega_{{\rm b},0})$,
$f_{{\rm b}}=\Omega_{{\rm b},0}/(\Omega_{{\rm c},0}+\Omega_{{\rm b},0})$,
$\Omega_{j,0}\equiv\bar{\rho}_{j,0}/\rho_{{\rm crit},0}$ is the present-day
mean density of component $j$ in the unit of the critical density,
$c_{s}$ is the sound speed, and $H$ is the Hubble constant at a
given redshift. Even though a usual approximation for $c_{s}$ is
a spatially uniform one given by

\begin{equation}
c_{s}^{2}=\frac{k_{B}\bar{T}}{\mu m_{{\rm H}}}\left(1-\frac{1}{3}\frac{\partial\log\bar{T}}{\partial\log a}\right),\label{eq:cs_simple}
\end{equation}
which assumes a mean-density environment undergoing Hubble expansion
with the mean baryon temperature $\bar{T}$, for high $k$ modes a
more accurate treatment is required (\citealt{Naoz2005}). This requires
replacing the pressure term%
\footnote{We keep the cross term $\delta_{{\rm b}}\delta_{T}$ 
in Equation (\ref{eq:mater_pressure})
in order to find the correct coupling of high-${\bf k}$ and low-${\bf k}$
modes in Equations (\ref{eq:perturbation_spatial}) and (\ref{eq:perturbation_k}).%
},

\begin{equation}
c_{s}^{2}\nabla\delta_{{\rm b}}\mapsto\frac{k_{B}\bar{T}}{\mu m_{{\rm H}}}\nabla\left(\delta_{{\rm b}}+\delta_{T}+\delta_{{\rm b}}\delta_{T}\right),\label{eq:mater_pressure}
\end{equation}
in Equation (\ref{eq:master}) and considering another rate equation

\begin{equation}
\frac{\partial\delta_{T}}{\partial t}=\frac{2}{3}\frac{\partial\delta_{{\rm b}}}{\partial t}+\frac{x_{e}(t)}{t_{\gamma}}a^{-4}\left\{ \delta_{T_{\gamma}}\left(\frac{5\bar{T}_{\gamma}}{\bar{T}}-4\right)-\delta_{T}\frac{\bar{T}_{\gamma}}{\bar{T}}\right\} \label{eq:master_temp}
\end{equation}
where $\delta_{T}$ and $\delta_{T_{\gamma}}$ are temperature fluctuations
of the baryon and the photon, respectively, $\bar{T}_{\gamma}\equiv2.725\,{\rm K}\,(1+z)$
is the mean photon temperature, $x_{e}(t)$ is the global electron
fraction at time $t$ and $t_{\gamma}\equiv1.17\times10^{12}\,{\rm yr}$.
For $\bar{T}$, we use the fitting formula by TH.

Equation (\ref{eq:master}) allows a trivial solution: $\delta_{{\rm c}}=\delta_{{\rm b}}=\phi=0$,
${\bf v}_{{\rm c}}={\bf v}_{{\rm c},i}(a/a_{i})^{-1}$ and ${\bf v}_{{\rm b}}={\bf v}_{{\rm b},i}(a/a_{i})^{-1}$,
where $a_{i}$ is the initial scale factor. TH took this as the zeroth-order
solution of a spatial patch and developed a linear perturbation theory
of small scale modes inside the patch. The physical process is in
principle a coupling of small-${\bf k}$ and large-${\bf k}$ modes
(mode-mode coupling), which is beyond the linear theory where all
modes are assumed to be mutually independent. TH used the fact that
the relative velocity ${\bf V}_{{\rm bc}}\equiv{\bf v}_{{\rm b}}-{\bf v}_{{\rm c}}$
is coherent over the length scale of a few comoving Mpc (contributed
by modes with wave numbers in the range $0.01\lesssim(k/{\rm Mpc}^{-1})\lesssim1$),
and averaged the ``local'' power spectra of density fluctuations
over many such patches with varying ${\bf V}_{{\rm bc}}$.

Even though a trivial solution exists, there also exists a nontrivial
solution to Equation (\ref{eq:master}), exact to the first order.
This nontrivial solution is suited to describe the physics inside
patches with non-zero overdensity (see also \citealt{Blazek2015}). In order to obtain the nontrivial
solution, we first linearize Equation (\ref{eq:master}):

\begin{eqnarray}
\frac{\partial\delta_{{\rm c}}}{\partial t} & = & -\theta_{{\rm c}},\nonumber \\
\frac{\partial\theta_{{\rm c}}}{\partial t} & = & -\frac{3}{2}H^{2}\Omega_{m}\left(f_{{\rm c}}\delta_{{\rm c}}+f_{{\rm b}}\delta_{{\rm b}}\right)-2H\theta_{{\rm c}},\nonumber \\
\frac{\partial\delta_{{\rm b}}}{\partial t} & = & -\theta_{{\rm b}},\nonumber \\
\frac{\partial\theta_{{\rm b}}}{\partial t} & = & -\frac{3}{2}H^{2}\Omega_{m}\left(f_{{\rm c}}\delta_{{\rm c}}+f_{{\rm b}}\delta_{{\rm b}}\right)-2H\theta_{{\rm b}},\label{eq:background}
\end{eqnarray}
where 
$\Omega_{m}\equiv\bar{\rho}_{m}(a)/\rho_{\rm crit}(a)$ is the
matter content with respect to the critical density $\rho_{\rm crit}$ at $a$, and
we ignored second-order terms and also the pressure term $a^{-1}c_{{\rm s}}^{2}\nabla\delta_{{\rm b}}$.
This is indeed a valid approximation in the wave number range ($0.01\lesssim(k/{\rm Mpc}^{-1})\lesssim1$)
relevant to the coherent ${\bf V}_{{\rm bc}}$ (TH), where the second-order
terms remain much smaller than the first-order terms and the baryonic
sound speed keeps decreasing from $c_{{\rm s}}\sim6\,$km/s after
recombination to make the pressure term negligible. Then, Equation
(\ref{eq:background}) can be rewritten as

\begin{eqnarray}
\frac{\partial\delta_{+}}{\partial t} & = & -\theta_{+},\nonumber \\
\frac{\partial\theta_{+}}{\partial t} & = & -\frac{3}{2}H^{2}\Omega_{m}\delta_{+}-2H\theta_{+},\nonumber \\
\frac{\partial\delta_{-}}{\partial t} & = & -\theta_{-},\nonumber \\
\frac{\partial\theta_{-}}{\partial t} & = & -2H\theta_{-},\label{eq:background_easy}
\end{eqnarray}
where $\delta_{+}\equiv f_{{\rm c}}\delta_{{\rm c}}+f_{{\rm b}}\delta_{{\rm b}}$,
$\theta_{+}\equiv f_{{\rm c}}\theta_{{\rm c}}+f_{{\rm b}}\theta_{{\rm b}}$,
$\delta_{-}\equiv\delta_{{\rm c}}-\delta_{{\rm b}}$, and $\theta_{-}\equiv\theta_{{\rm c}}-\theta_{{\rm b}}$.
In the matter-dominated ($\Omega_{m}=1$) flat universe, $\delta_{+}$ allows both the
growing mode ($\delta_{+}\propto a$, $\theta_{+}\propto a^{-1/2}$,
${\bf v}_{+}\equiv f_{{\rm c}}{\bf v}_{{\rm c}}+f_{{\rm b}}{\bf v}_{{\rm b}}\propto a^{1/2}$)
and the decaying mode ($\delta_{+}\propto a^{-3/2}$, $\theta_{+}\propto a^{-3}$,
${\bf v}_{+}\propto a^{-2}$). $\delta_{-}$ allows a slowly decaying (``streaming'')
mode ($\delta_{-}\propto a^{-1/2}$, $\theta_{-}\propto a^{-2}$,
${\bf v}_{-}\equiv{\bf v}_{{\rm c}}-{\bf v}_{{\rm b}}\propto a^{-1}$)
and a compensated mode ($\delta_{-}$=constant, $\theta_{-}=0$, ${\bf v}_{-}=0$).
During $1000\gtrsim z\gtrsim 50$, the non-negligible amount of
  the radiation component (CMB
  and neutrinos) makes $\Omega_{m} \ne 1$, and most of the simple analytical forms
above become no longer intact except for \{$\theta_{-}$, ${\bf v}_{-}$\} of the streaming
mode and \{$\delta_{-}$, $\theta_{-}$, ${\bf v}_{-}$\} of the compensated
mode.

Using this mode decomposition, the large-scale perturbations
  evolve in the following form:
\begin{widetext}
\begin{eqnarray}
\Delta_{{\rm c}}(a) & = & \left\{\Delta_{\rm gro}D^{\rm g}(a)+\Delta_{\rm dec}D^{\rm d}(a)\right\}+f_{{\rm b}}\left\{\Delta_{\rm com}+\Delta_{\rm str}D^{\rm s}(a)\right\},\nonumber \\
\Delta_{{\rm b}}(a) & = & \left\{\Delta_{\rm gro}D^{\rm g}(a)+\Delta_{\rm dec}D^{\rm d}(a)\right\}-f_{{\rm c}}\left\{\Delta_{\rm com}+\Delta_{\rm str}D^{\rm s}(a)\right\},\nonumber \\
\Theta_{{\rm c}}(a) & = & -a H \left\{\Delta_{\rm gro}\frac{dD^{\rm g}(a)}{da}+\Delta_{\rm dec}\frac{dD^{\rm d}(a)}{da}\right\}+f_{{\rm b}}\left(\Theta_{{\rm c},i}-\Theta_{{\rm b},i}\right)\left(\frac{a}{a_{i}}\right)^{-2},\nonumber \\
\Theta_{{\rm b}}(a) & = & -a H \left\{\Delta_{\rm gro}\frac{dD^{\rm g}(a)}{da}+\Delta_{\rm dec}\frac{dD^{\rm d}(a)}{da}\right\}-f_{{\rm c}}\left(\Theta_{{\rm c},i}-\Theta_{{\rm b},i}\right)\left(\frac{a}{a_{i}}\right)^{-2},\nonumber \\
{\bf V}_{{\rm bc}}(a) & = & \left({\bf V}_{{\rm c},i}-{\bf V}_{{\rm b},i}\right)\left(\frac{a}{a_{i}}\right)^{-1},\label{eq:background_sol}
\end{eqnarray}
\end{widetext}
where we used upper-case letters to denote the ``background'' fluctuations
for each patch of a few comoving Mpc, over which we will develop the
small-scale perturbation. 
$\Delta_{\rm gro}$, $\Delta_{\rm dec}$,
$\Delta_{\rm com}$, and $\Delta_{\rm str}$ are the initial (at
$z=1000$) values of the growing, decaying, compensated, and streaming
modes, respectively. $D^{\rm g}(a)$, $D^{\rm d}(a)$, and $D^{\rm
  s}(a)$ are growth factors of the growing, decaying,
and streaming modes, respectively, and they are all normalized
as $D^{\rm g}=D^{\rm d}=D^{\rm s}=1$ at $z=1000$. We describe the
details of these modes in Appendix A.
It is noteworthy, as is
well known already, that both CDM and baryonic components tend
  to approach
the same asymptotes $\Delta=\Delta_{\rm gro}D^{\rm g}(a)$
and $\Theta=-aH\Delta_{\rm gro}dD^{\rm g}(a)/da$,
indicating that baryons tend to move together with CDMs in time. More
interestingly, ${\bf V}_{{\rm bc}}$ decays as $a^{-1}$ throughout
the evolution at any overdensity environment even though ${\bf V}_{{\rm c}}$
and ${\bf V}_{{\rm b}}$ grow roughly as $a^{1/2}$ individually (except in
regions with $\Delta_{{\rm c}}=\Delta_{{\rm b}}=0$ where ${\bf V}_{{\rm c}}$
and ${\bf V}_{{\rm b}}$ decay as $a^{-1}$). This fact may seem to
make the analysis by TH valid in generic overdensity environments
to some extent: TH relied on the trivial solution $\Delta_{{\rm c}}=\Delta_{{\rm b}}=0$,
in which all velocity components decay in time such that ${\bf V}_{{\rm c}}\propto a^{-1}$,
${\bf V}_{{\rm b}}\propto a^{-1}$, and most importantly ${\bf V}_{{\rm bc}}\propto a^{-1}$.
Because the suppression of the matter-density fluctuations ($f_{{\rm c}}\delta_{{\rm c}}+f_{{\rm b}}\delta_{{\rm b}}$)
depends not on individual velocity components but on ${\bf V}_{{\rm bc}}$
only, different temporal behavior of individual velocity components
among the trivial and generic solutions do not matter. Nevertheless,
quantitative prediction by TH will be questioned in Section \ref{sub:Pk},
because we use nontrivial solutions (Equation \ref{eq:background_sol})
which result in the new type of coupling of high-${\bf k}$ and low-${\bf k}$
modes in general. We also require the evolution of $\Delta_{T}$,
which is given by Equation (\ref{eq:master_temp}):

\begin{equation}
\frac{\partial\Delta_{T}}{\partial
  t}=\frac{2}{3}\frac{\partial\Delta_{{\rm b}}}{\partial
  t}-\frac{x_{e}(t)}{t_{\gamma}}a^{-4}\frac{\bar{T}_{\gamma}}{\bar{T}}\Delta_{T},
\label{eq:Deltatemp}
\end{equation}
which is evolved in conjunction with Equation (\ref{eq:master}).
Here we neglect $\Delta_{T_\gamma}$ term due to its smallness (see
also the following discussion), even though we do not neglect
$\Delta_{T_\gamma}(a_{i})$ when initializing $\Delta_{T}(a_{i})$ (Section
\ref{sub:numerical}).
We have found a useful fitting formula for $\Delta_{T}(a)$ for patches
with volume $(4\,{\rm Mpc})^{3}$:

\begin{eqnarray}
\Delta_{T}(a)&=&{\rm sign}(\Delta_{T,\, A})\,{\rm dex}\left[\alpha\left(\log_{10}(a)+2.8\right)^{0.33}\right]\nonumber \\
&&|\Delta_{T,\, A}|^{4.2591}|\Delta_{T,\, B}|^{-3.2591},
\label{eq:DeltaT_fitting}
\end{eqnarray}
which provides a good fit to $\Delta_{T}$ at $300\gtrsim z\gtrsim 5$, and
for higher redshift range we simply ignore $\Delta_{T}$ altogether
because of smallness of $\Delta_{T}$ in general. Here $\Delta_{T,\,
  A}\equiv\Delta_{T}(a=0.01)=0.279\Delta_{{\rm b}}(a=0.01)$, 
$\alpha\equiv\log_{10}(\Delta_{T,\, B}/\Delta_{T,\, A})/0.28505$
and $\Delta_{T,\, B}\equiv\Delta_{T}(a=0.1)=0.599\Delta_{{\rm b}}(a=0.1)$,
and an almost complete coupling of $\Delta_{T}(a)$ to $\Delta_{{\rm b}}(a)$
at $z\lesssim300$ yields this simple, empirical relation to $\Delta_{{\rm b}}(a)$.
We describe this fitting formula in more details in Appendix B.
The decoupling of $\Delta_{T}(a)$ from the CMB is much earlier than
the mean value experiences, which occur at $z\simeq150$, because
the larger the ${\bf k}$ is, the earlier the decoupling occurs (see
e.g. Figure 1 of \citealt{Naoz2005}). Including $\Delta_{T_\gamma}$
explicitly may delay this decoupling to some extent, but we leave such
an accurate calculation to future work.
Using the evolution equation for $\Delta_{{\rm b}}(a)$
in Equation (\ref{eq:background_sol}), Equation (\ref{eq:DeltaT_fitting})
is determined solely by local values of the 4 modes.

Now we expand Equations (\ref{eq:master}), (\ref{eq:mater_pressure})
and (\ref{eq:master_temp}) to the linear order, taking Equation (\ref{eq:background_sol})
as the zeroth-order solution. We first define the net density, the
net velocity (of the fluid component $i$), the net gravitational
potential and the net baryon temperature as $\rho_{i}(a,\,{\bf x})=\bar{\rho}_{i}(a)\{1+\Delta_{i}(a,\,{\bf X})+\delta_{i}(a,\,{\bf x})\}$,
${\bf v}_{i,{\rm net}}(a,\,{\bf x})=Ha{\bf x}+{\bf V}_{i}(a,\,{\bf X})+{\bf v}_{i}(a,\,{\bf x})$,
$\phi_{{\rm net}}(a,\,{\bf x})=-(a/2)\partial(Ha)/\partial t+\Phi(a,\,{\bf X})+\phi(a,\,{\bf x})$,
and $T_{{\rm b},{\rm net}}=\bar{T}(1+\Delta_{T}+\delta_{T})$, respectively.
Here ${\bf X}$ and ${\bf x}$ denote the comoving-coordinate position
of the center of a background patch and that of a small-scale
fluid component, respectively, and we use lower-case letters for small-scale
fluctuations. $\Phi$ is the gravitational potential sourced only
by the background fluctuations such that $\nabla^{2}\Phi=4\pi Ga^{2}\Omega_{m}(f_{{\rm c}}\Delta_{{\rm c}}+f_{{\rm b}}\Delta_{{\rm b}})$
(e.g. \citealt{Bernardeau2002}). Similarly, $\nabla^{2}\phi=4\pi Ga^{2}\Omega_{m}(f_{{\rm c}}\delta_{{\rm c}}+f_{{\rm b}}\delta_{{\rm b}})$.
We then have

\begin{figure}
\includegraphics[width=0.5\textwidth]{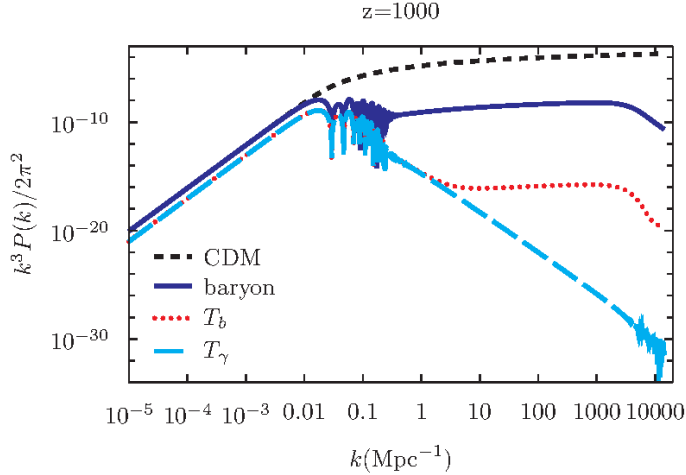}

\caption{Fluctuations of the CDM density (short-dashed, black), the baryon
density (solid, blue), the baryon temperature (dotted, red) and the
photon temperature (long-dashed, cyan), represented by the ${\bf k}$-space
variance at $z=1000$.}
\label{fig:Pk_z1000}

\end{figure}

\begin{widetext}
\begin{eqnarray}
\frac{\partial\delta_{{\rm c}}}{\partial t} & = & -a^{-1}{\bf V}_{{\rm c}}\cdot\nabla\delta_{{\rm c}}-(1+\Delta_{{\rm c}})\theta_{{\rm c}}-\Theta_{{\rm c}}\delta_{{\rm c}},\nonumber \\
\frac{\partial{\bf v}_{{\rm c}}}{\partial t} & = & -a^{-1}\left({\bf V}_{{\rm c}}\cdot\nabla\right){\bf v}_{{\rm c}}-a^{-1}\nabla\phi-H{\bf v}_{{\rm c}}-\left[a^{-1}\left({\bf v}_{{\rm c}}\cdot\nabla\right){\bf V}_{{\rm c}}\right],\nonumber \\
\frac{\partial\delta_{{\rm b}}}{\partial t} & = & -a^{-1}{\bf V}_{{\rm b}}\cdot\nabla\delta_{{\rm b}}-(1+\Delta_{{\rm b}})\theta_{{\rm b}}-\Theta_{{\rm b}}\delta_{{\rm b}},\nonumber \\
\frac{\partial{\bf v}_{{\rm b}}}{\partial t} & = & -a^{-1}\left({\bf V}_{{\rm b}}\cdot\nabla\right){\bf v}_{{\rm b}}-a^{-1}\nabla\phi-H{\bf v}_{{\rm b}}\nonumber -a^{-1}\frac{k_{B}\bar{T}}{\mu m_{{\rm H}}}\nabla\left\{ \left(1+\Delta_{{\rm b}}\right)\delta_{T}+\left(1+\Delta_{T}\right)\delta_{{\rm b}}\right\} -\left[a^{-1}\left({\bf v}_{{\rm b}}\cdot\nabla\right){\bf V}_{{\rm b}}\right],\nonumber \\
\frac{\partial\delta_{T}}{\partial t} & = & \frac{2}{3}\left\{ \frac{\partial\delta_{{\rm b}}}{\partial t}+\frac{\partial\Delta_{{\rm b}}}{\partial t}\left(\delta_{T}-\delta_{{\rm b}}\right)+\frac{\partial\delta_{{\rm b}}}{\partial t}\left(\Delta_{T}-\Delta_{{\rm b}}\right)\right\} \nonumber +\frac{x_{e}(t)}{t_{\gamma}}a^{-4}\left\{ \left[\left(\frac{5\bar{T}_{\gamma}}{\bar{T}}-4\right)\delta_{T_{\gamma}}\right]-\frac{\bar{T}_{\gamma}}{\bar{T}}\delta_{T}+4\left[\Delta_{T_{\gamma}}\delta_{T}+\Delta_{T}\delta_{T_{\gamma}}\right]\right\} \nonumber \\
\nabla^{2}\phi & = & 4\pi Ga^{2}\Omega_{m}(f_{{\rm c}}\delta_{{\rm c}}+f_{{\rm b}}\delta_{{\rm b}}),\label{eq:perturbation_spatial}
\end{eqnarray}
\end{widetext}
where we marked the terms that can be further ignored in square brackets.
First, we can safely ignore any terms containing $\delta_{T_{\gamma}}$
and $\Delta_{T_{\gamma}}$, because the former is negligible at $z\lesssim1000$
compared to $\delta_{T}$ (see Figure \ref{fig:Pk_z1000}) and the
latter is just too small ($\Delta_{T_{\gamma}}\lesssim10^{-5}$ at
$z=1000$ and decaying in time) to produce any appreciable
impact on the baryon temperature of high ${\bf k}$ modes. Secondly,
the justification for ignoring $a^{-1}\left({\bf v}_{i}\cdot\nabla\right){\bf V}_{i}$
is easily seen in viewpoint of the ${\bf k}$-space.
With the Fourier expansion $A({\bf x})=\sum_{{\bf k}}A({\bf k})\exp(i{\bf k}\cdot{\bf x})$
of a quantity $A({\bf x})$ in an actual, real space (``${\bf r}$
space'' henceforth), the background velocities have ${\bf V}_{j}({\bf x})=\sum_{{\bf K}}{\bf V}_{j}({\bf K})\exp(i{\bf K}\cdot{\bf x})$
but with the condition ${\bf K}\lesssim1\,{\rm Mpc}^{-1}$. In contrast,
the small-scale modes fluctuating against the background patches have
intrinsically larger wave number ${\bf k}$, or $K\ll k$. Then, at
each ${\bf K}$, $\left|\left({\bf v}_{j}\cdot\nabla\right){\bf V}_{j}({\bf K})\right|\sim Kv_{j}V_{j}\ll\left|\left({\bf V}_{j}({\bf K})\cdot\nabla\right){\bf v}_{j}\right|\sim kv_{j}V_{j}$.
Similarly, $\left|\nabla\Delta_{{\rm b}}\right|\sim K\Delta_{{\rm b}}\ll\left|\nabla\delta_{{\rm b}}\right|\sim k\delta_{{\rm b}}$.
We finally note that we do not include the coupling terms between
  similar wavenumbers in Equation (\ref{eq:perturbation_spatial}), which
  will involve quadratic and higher-order
  polynomicals of $\Delta$ and $\delta$. Therefore, the validity of Equation
  (\ref{eq:perturbation_spatial}) will break down in the non-linear
  regime. Nevertheless, our approach is more suitable for a crude
  estimattion of the conditional halo mass function (Section
  \ref{sub:mass_function}) in terms of the
  extended Press-Schechter formalism, which is based on the mapping 
  of the linear density growth to the nonlinear growth of the
  e.g. top-hat density perturbation.

The evolution of small-scale perturbations, therefore, is coupled
to large-scale perturbations on which they are sitting. Ignoring the
terms in square brackets, and shifting the viewpoint to the CDM rest
frame in which ${\bf V}_{{\rm c}}=0$ (as in \citealt{O'Leary2012};
TH chose the baryon rest frame), Equation (\ref{eq:perturbation_spatial}),
in the ${\bf k}$-space, finally becomes

\begin{eqnarray}
\frac{\partial\delta_{{\rm c}}}{\partial t} & = & -(1+\Delta_{{\rm c}})\theta_{{\rm c}}-\Theta_{{\rm c}}\delta_{{\rm c}},\nonumber \\
\frac{\partial\theta_{{\rm c}}}{\partial t} & = & -\frac{3}{2}H^{2}\Omega_{m}\left(f_{{\rm c}}\delta_{{\rm c}}+f_{{\rm b}}\delta_{{\rm b}}\right)-2H\theta_{{\rm c}},\nonumber \\
\frac{\partial\delta_{{\rm b}}}{\partial t} & = & -ia^{-1}{\bf V}_{{\rm bc}}\cdot{\bf k}\delta_{{\rm b}}-(1+\Delta_{{\rm b}})\theta_{{\rm b}}-\Theta_{{\rm b}}\delta_{{\rm b}},\nonumber \\
\frac{\partial\theta_{{\rm b}}}{\partial t} & = & -ia^{-1}{\bf V}_{{\rm bc}}\cdot{\bf k}\theta_{{\rm b}}-\frac{3}{2}H^{2}\Omega_{m}\left(f_{{\rm c}}\delta_{{\rm c}}+f_{{\rm b}}\delta_{{\rm b}}\right)-2H\theta_{{\rm b}} \nonumber \\
&&+a^{-2}\frac{k_{B}\bar{T}}{\mu m_{{\rm H}}}k^{2}\left\{ \left(1+\Delta_{{\rm b}}\right)\delta_{T}+\left(1+\Delta_{T}\right)\delta_{{\rm b}}\right\} ,\nonumber \\
\frac{\partial\delta_{T}}{\partial t} & = & \frac{2}{3}\left\{
\frac{\partial\delta_{{\rm b}}}{\partial t}+\frac{\partial\Delta_{{\rm b}}}{\partial t}\left(\delta_{T}-\delta_{{\rm b}}\right)+\frac{\partial\delta_{{\rm b}}}{\partial t}\left(\Delta_{T}-\Delta_{{\rm b}}\right)\right\} \nonumber \\
&&-\frac{x_{e}(t)}{t_{\gamma}}a^{-4}\frac{\bar{T}_{\gamma}}{\bar{T}}\delta_{T},\label{eq:perturbation_k}
\end{eqnarray}
where $\delta_{j}$, $\theta_{j}$ and $\delta_{T}$ now denote fluctuations
in the ${\bf k}$-space while $\Delta_{j}$, $\Theta_{j}$ and ${\bf V}_{j}$
are fluctuations of a given patch at $(a,\,{\bf X})$ in the ${\bf r}$
space, given by Equation (\ref{eq:background_sol}).

\subsection{Evolution of perturbation inside patches: Numerical Scheme}
\label{sub:numerical}

Evolution of small-scale perturbations can be calculated by integrating
the rate equation (Equation \ref{eq:perturbation_k}) from some initial
redshift, preferentially not too long after the recombination epoch
when the relative motion has not yet influenced the evolution. We
take $z_{i}\equiv1000$ as the initial redshift. The initial condition
should be generated for both the background quantities and the small-scale
modes. For the background, as perturbations in Equation (\ref{eq:perturbation_k})
are ${\bf r}$-space quantities whose distributions are all Gaussian,
one needs to sample these values in the ${\bf r}$-space accordingly.
For the small-scale modes, one just needs to track the evolution of
the average value in the ${\bf k}$-space.

Let us first describe the statistics of background patches that we
expect. TH calculated the evolution of small-scale ($k\gtrsim10$)
fluctuations under different background patches but only of $\Delta_{{\rm c}}=\Delta_{{\rm b}}=0$,
and defined ``local power spectrum'' $P_{{\rm loc,m}}(k;\, V{}_{{\rm bc}})$
averaged out over all possible opening angles between ${\bf V}_{{\rm bc}}$
and ${\bf k}$. In our case, there are extra dimensions to consider
which are $\Delta_{{\rm c}}$, $\Delta_{{\rm b}}$, $\Theta_{{\rm c}}$,
and $\Theta_{{\rm b}}$, resulting in a much higher computational
demand. Fortunately, some of these quantities are in perfect correlation
with one another with linear proportionality. In addition, their initial
values at $a=a_{i}$ completely compose the ensemble at any time through
Equation (\ref{eq:background_sol}). At the minimal level\footnote{For a more accurate treatment or for a specific patch of interest,
one should also consider variations in other variables, such that
$P_{{\rm loc,m}}=P_{{\rm loc,m}}\left(k;\, V_{{\rm bc}};\,\Delta_{{\rm c}};\,\Delta_{{\rm b}};\,\Theta_{{\rm c}};\,\Theta_{{\rm b}};\,\Delta_{T};\,\Delta_{T_\gamma}\right)$.
As the correlation between $V_{{\rm b}}$ ($\Delta_{{\rm b}}$) and
$V_{{\rm c}}$ ($\Delta_{{\rm c}}$) becomes tighter in time, the
initial variation gets gradually diluted, which roughly justifies
our restricting the parameter space only to $V_{{\rm bc}}$ and $\Delta_{{\rm c}}$.%
}, it would
suffice to just consider variation of $\Delta_{{\rm c}}$ in addition
to $V_{{\rm bc}}$ such that the local power spectrum is an explicit
function of the two background quantities, or $P_{{\rm loc,m}}=P_{{\rm loc,m}}\left(k;\, V_{{\rm bc}}(a_{i});\,\Delta_{{\rm c}}(a_{i})\right)$.

For the initial condition for background patches, we generate 3D maps
of $\Delta_{{\rm c}}$, $\Delta_{{\rm b}}$, $\Theta_{{\rm c}}$,
$\Theta_{{\rm b}}$, ${\bf V}_{{\rm c}}$, ${\bf V}_{{\rm b}}$, and
$\Delta_{T}$ at $z=1000$ on $151^3$ uniform grid cells
inside a cubical volume of $V_{{\rm box}}=(604\,{\rm Mpc})^{3}$.
We generate fluctuations of discrete modes that are randomized as

\begin{eqnarray}
{\rm Re}(\Delta_{{\bf k}}) & = & G_{1}N^{3}\left(\frac{P(k)}{2V_{{\rm box}}}\right)^{1/2}{\rm sign}\left[{\tt TF}(\Delta_{{\bf k}})\right],\nonumber \\
{\rm Im}(\Delta_{{\bf k}}) & = & G_{2}N^{3}\left(\frac{P(k)}{2V_{{\rm box}}}\right)^{1/2}{\rm sign}\left[{\tt TF}(\Delta_{{\bf k}})\right],\label{eq:random}
\end{eqnarray}
for given ${\bf k}$, where $G_{1}$ and $G_{2}$ are random numbers
drawn from mutually independent Gaussian distributions with mean 0
and standard deviation 1, $\Delta_{{\bf k}}$ stands for any kind
of ${\bf k}$-space fluctuations, and \texttt{TF} is the transfer
function of $\Delta_{{\bf k}}$, whose sign should be multiplied because
some $\Delta_{{\bf k}}$'s oscillate around zero in ${\bf k}$. The
configuration is roughly equivalent to applying a smoothing filter
of length $(604/151)=4$ Mpc. In practice, we use CAMB (\citealt{Lewis2000})
for $\Delta_{j}({\bf k},\, a_{i})$, 
and use the continuity equations ($\partial\Delta_{j}/\partial t=-\Theta_{j}$)
for $\Theta_{j}({\bf k},\, a_{i})$, with the help of two CAMB transfer-function
outputs at mutually nearby redshifts for time differentiation. ${\bf V}_{j}({\bf k},\, a_{i})$
is obtained from the relation ${\bf V}_{j}=-(ia{\bf k}/k^{2})\Theta_{j}$.
$\Delta_{T}({\bf k},\, a_{i})$ is fixed by following the scheme by
\citet{Naoz2005}: we require $\partial\Delta_{T}/\partial t=\partial\Delta_{T_{\gamma}}/\partial t$
at the initial redshift in Equation (\ref{eq:master_temp}), which
results in

\begin{equation}
\Delta_{T}=\Delta_{T_{\gamma}}\left(5-\frac{4\bar{T}}{\bar{T}_{\gamma}}\right)+\frac{t_{\gamma}}{x_{e}(t_{i})}a_{i}^{4}\left(\frac{2}{3}\frac{\partial\Delta_{{\rm b}}}{\partial t}-\frac{\partial\Delta_{T_{\gamma}}}{\partial t}\right)\label{eq:DeltaT_init}
\end{equation}
where all quantities are evaluated at $a_{i}$, especially with the
help of Equation (\ref{eq:background_sol}) for $\partial\Delta_{{\rm b}}({\bf k},\, a_{i})/\partial t$
and two adjacent CAMB transfer-function outputs for $\partial\Delta_{T_{\gamma}}({\bf k},\, a_{i})/\partial t=(1/4)\partial\Delta_{\gamma}({\bf k},\, a_{i})/\partial t$.
Finally, all these ${\bf k}$-space fluctuations are Fourier-transformed
to obtain ${\bf r}$-space fluctuations.

\begin{figure*}
\gridline{
\fig{fig2a}{0.28\textwidth}{(a)}
\fig{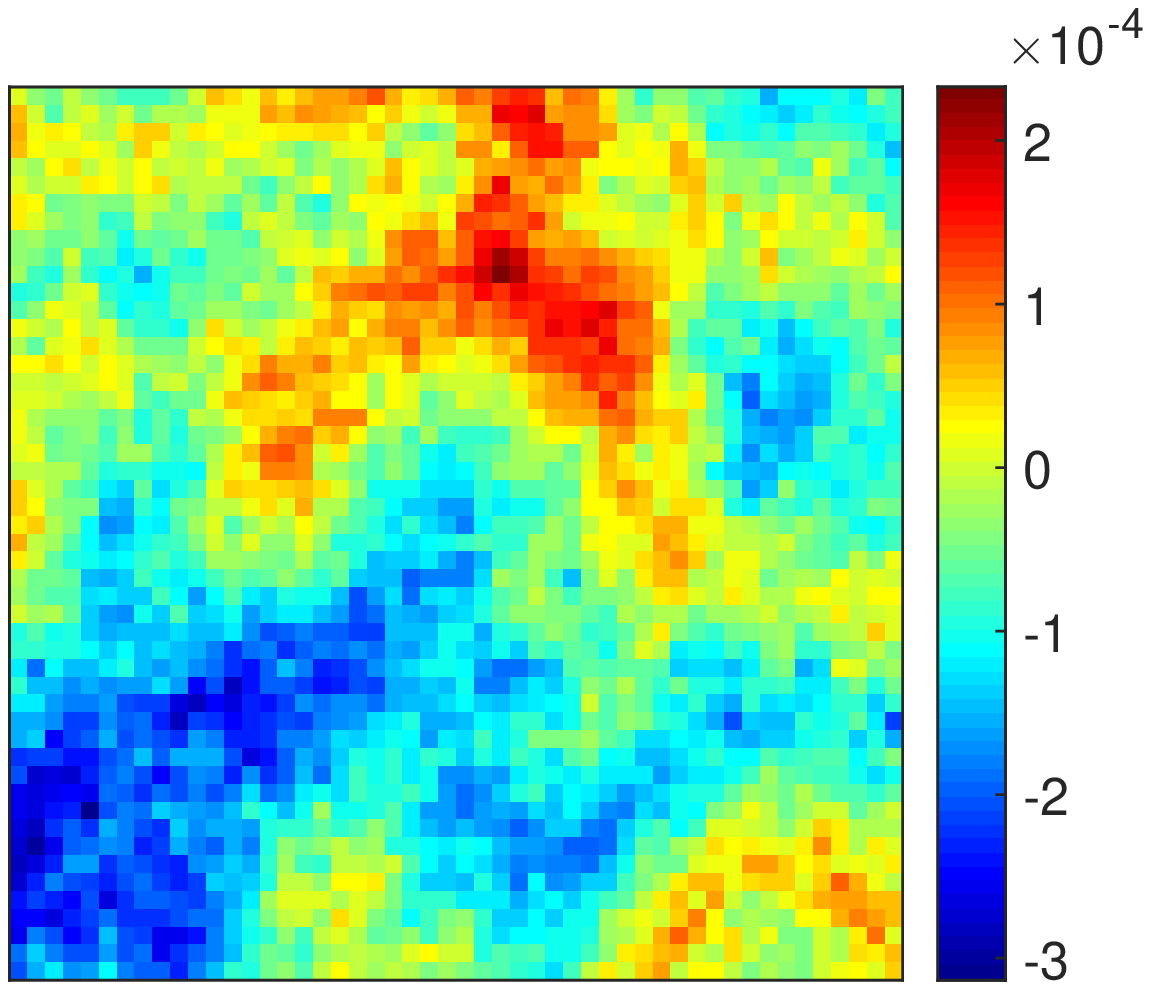}{0.28\textwidth}{(b)}
\fig{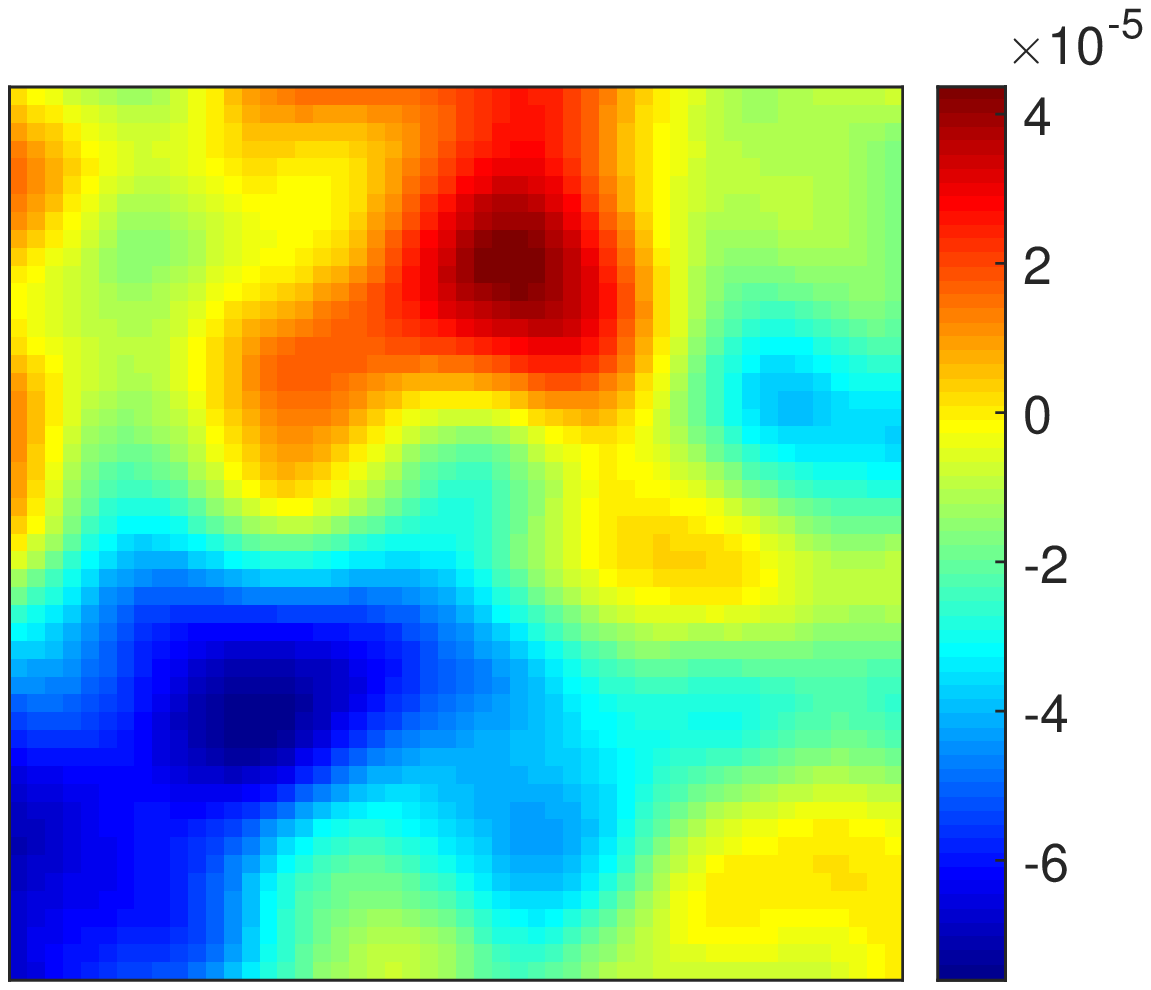}{0.28\textwidth}{(c)}
}

\gridline{
\fig{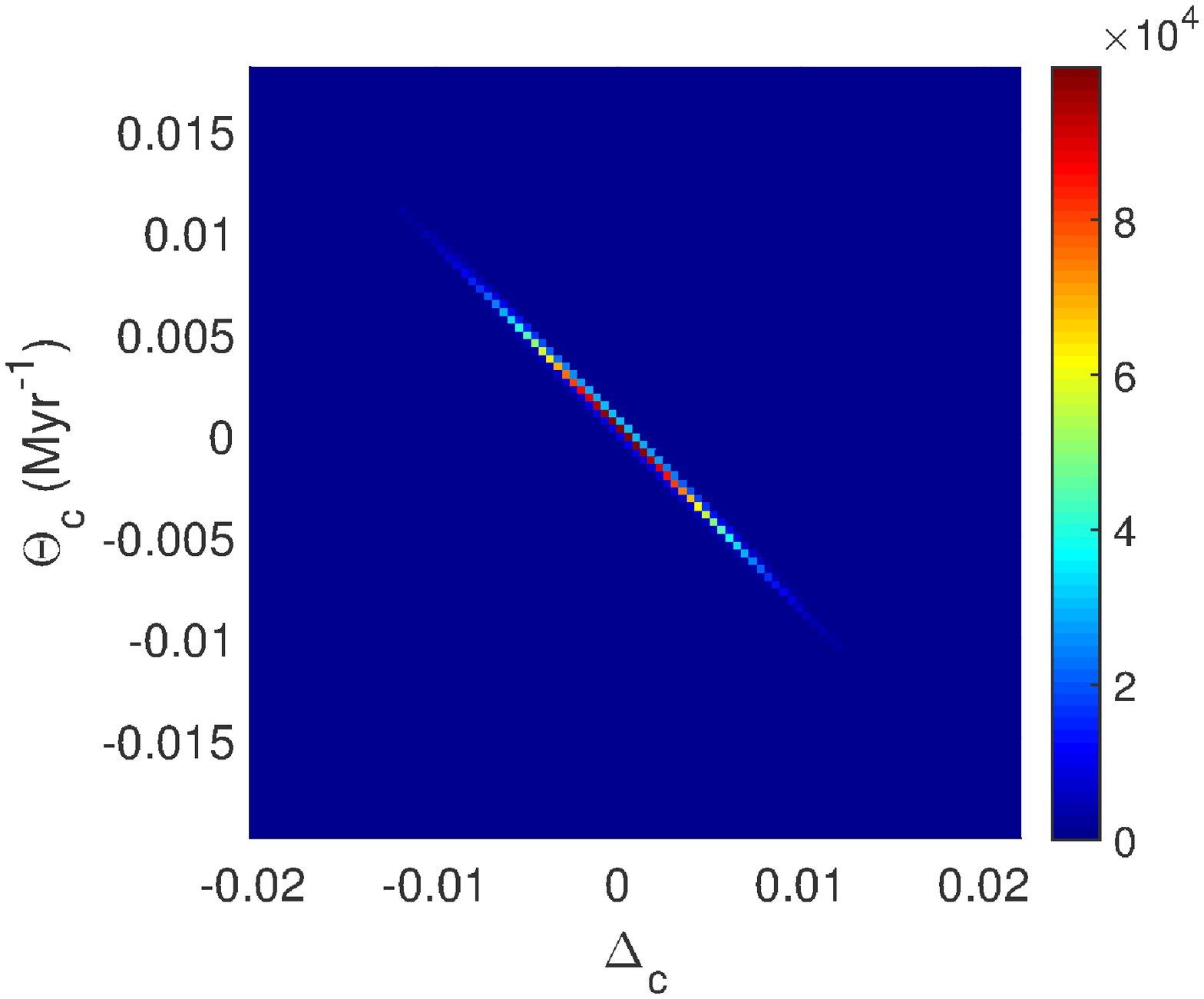}{0.33\textwidth}{(d)}
\fig{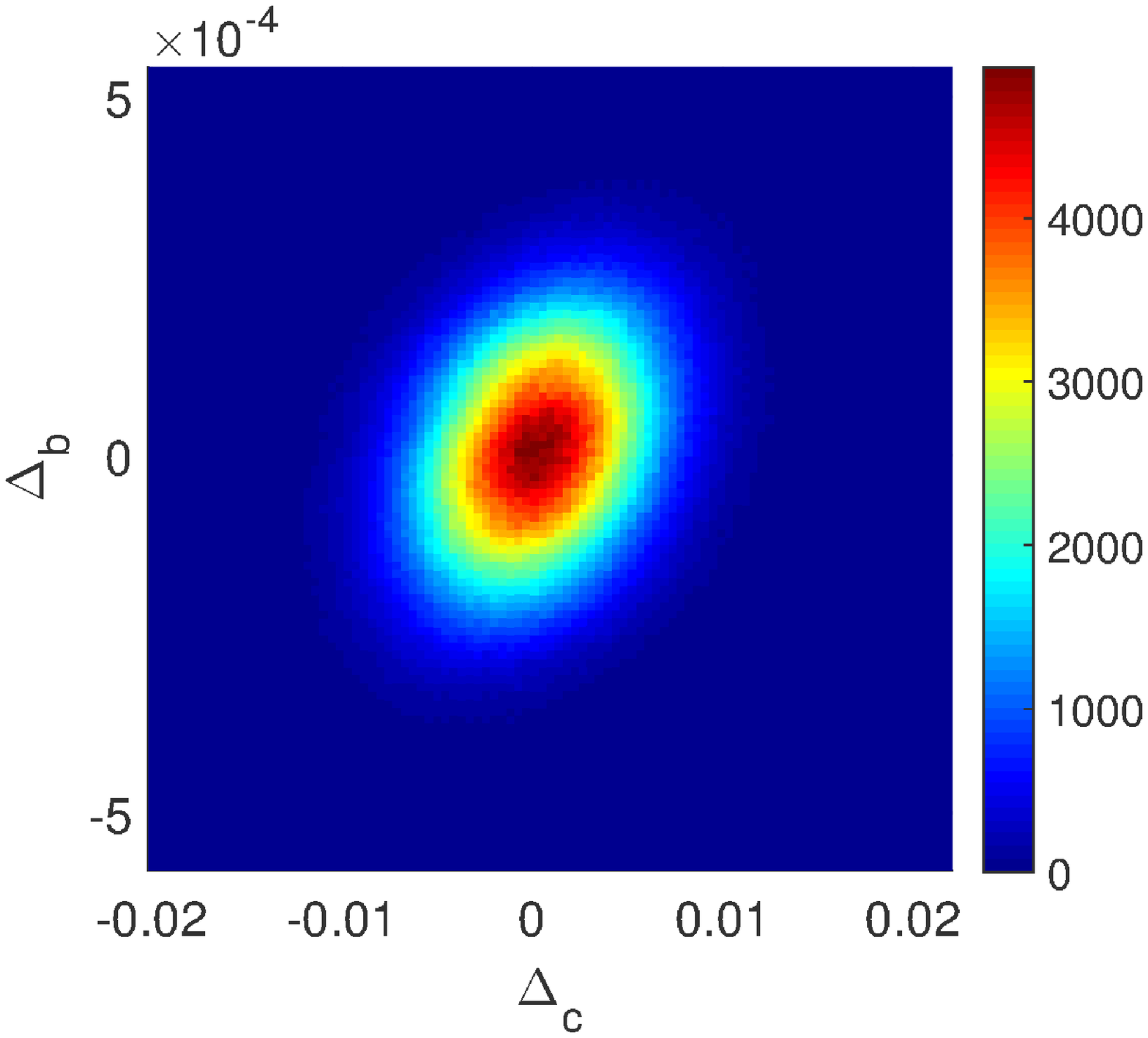}{0.33\textwidth}{(e)}
\fig{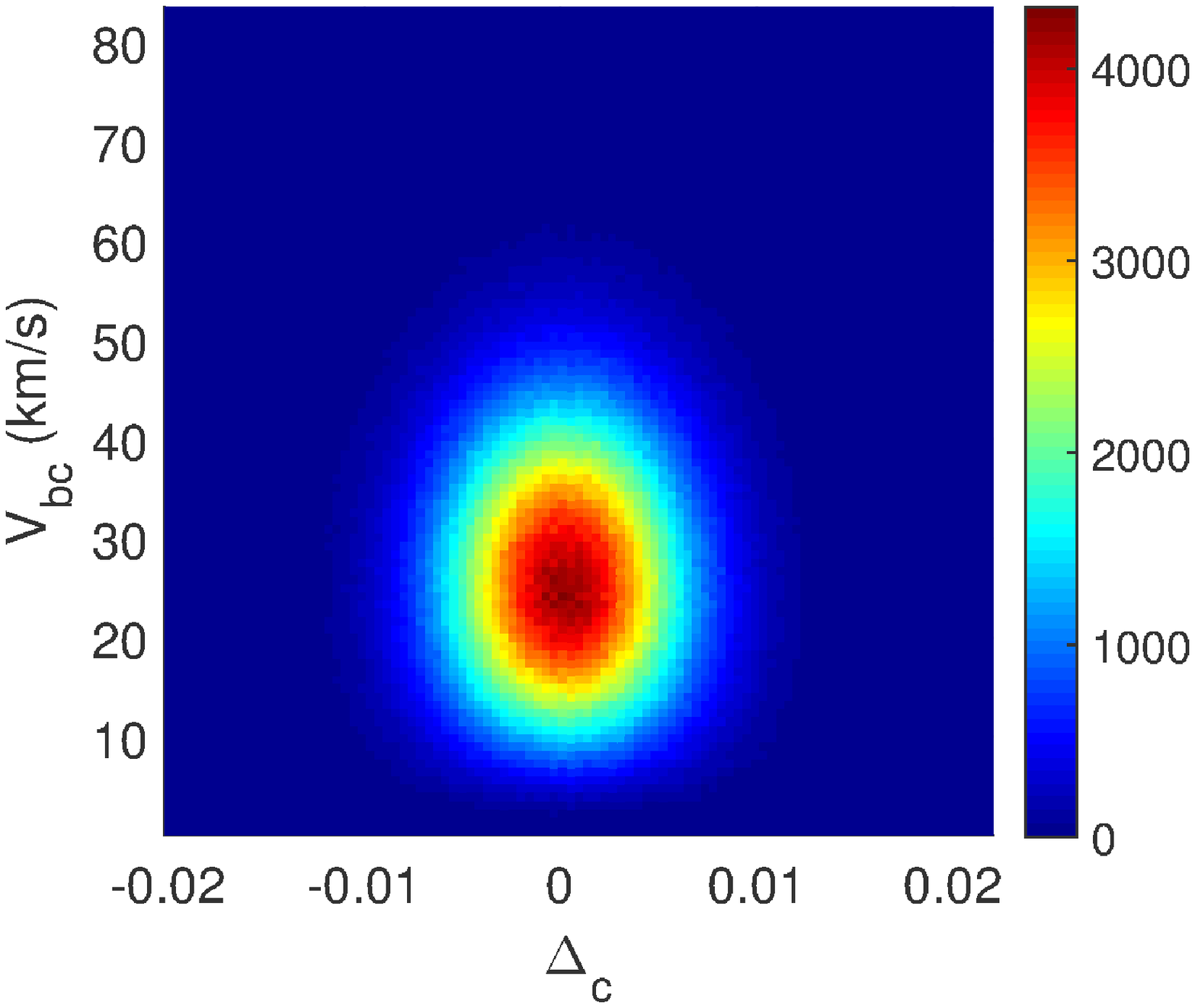}{0.33\textwidth}{(f)}
}

\caption{(a) 2D map of CDM overdensity $\Delta_{{\rm c}}$ (colored cells)
and relative velocity ${\bf V}_{{\rm cb}}\equiv -{\bf V}_{{\rm bc}}={\bf V}_{{\rm c}}-{\bf V}_{{\rm b}}$
(arrows; projected on the plane) fields on a slice of $200^{2}$ Mpc$^{2}$
containing $50^{2}$ cells. The plotted slice is an arbitrarily chosen
part of the actual volume of $604^{3}$ Mpc$^{3}$ we used, containing
$151^{3}$ cells in total. (b) 2D map of baryon overdensity $\Delta_{{\rm b}}$
on the same slice. (c) 2D map of baryon temperature overdensity $\Delta_{T}$
on the same slice. (d) Distribution of $\Delta_{{\rm c}}$ and CDM
velocity divergence $\Theta_{{\rm c}}$ (in units of Myr$^{-1}$).
The color bar represents the number of cells in sampling bins. (e)
Distribution of $\Delta_{{\rm c}}$ (x-axis) and $\Delta_{{\rm b}}$
(y-axis). (f) Distribution of $\Delta_{{\rm c}}$ and the relative
velocity $V_{{\rm bc}}=\left|{\bf V}_{{\rm bc}}\right|$ (in units
of km/s). All figures use quantities at $z=1000$.}
\label{fig:map}
\end{figure*}

3D maps and 2D histograms of several initial quantities are presented
in Figure \ref{fig:map}. Fields of $\Delta_{{\rm c}}$ and ${\bf V}_{{\rm bc}}$
on a part of a slice of the box at $z=z_{i}=1000$ are shown in Figure \ref{fig:map}(a).
As expected, the velocity field converges on overdense regions and
diverges on underdense regions. We find that in most patches
  $V_{{\rm c}}$ dominates over $V_{{\rm b}}$, and thus the map
of ${\bf V}_{{\rm c}}$ looks very similar
to Figure \ref{fig:map}(a). This occurs because baryons lag behind
CDMs due to their coupling to CMB. $\Delta_{{\rm b}}$ (Figure \ref{fig:map}b)
is coupled to $\Delta_{T}$ (Figure \ref{fig:map}c) more strongly than to $\Delta_{{\rm c}}$. $\Delta_{{\rm c}}$ and $\Theta_{{\rm c}}$
(similarly $\Delta_{{\rm b}}$ and $\Theta_{{\rm b}}$) are almost
perfectly correlated (Figure \ref{fig:map}d). $\Delta_{{\rm c}}$ and
$\Delta_{{\rm b}}$ are very loosely correlated due to the tight coupling
of baryons to photons at the redshift (Figure \ref{fig:map}e), but the
correlation becomes tighter in time. $\Delta_{{\rm c}}$
and $V_{{\rm bc}}$ are not correlated (Figure \ref{fig:map}f). Because
of this fact, the probability distribution function (PDF) $\mathcal{P}$
is simply a multiplication of PDFs $\mathcal{P}(k;\, V_{{\rm bc}})$
and $\mathcal{P}(k;\,\Delta_{{\rm c}})$, at the minimal level. Due to Gaussianity at $z_{i}$, we have

\begin{eqnarray}
\mathcal{P}(k;\,\Delta_{{\rm c}}) & = & \frac{1}{\sqrt{2\pi}\sigma_{\Delta_{{\rm c}}}}\exp\left[-\frac{\Delta_{{\rm c}}^{2}}{2\sigma_{\Delta_{{\rm c}}}^{2}}\right],\nonumber \\
\mathcal{P}(k;\, V_{{\rm bc}}) & = & \sqrt{\frac{2}{\pi}}\frac{V_{{\rm bc}}^{2}}{\sigma_{V_{{\rm bc}}}^{3}}\exp\left[-\frac{V_{{\rm bc}}^{2}}{2\sigma_{V_{{\rm bc}}}^{2}}\right],\label{eq:pdf}
\end{eqnarray}
where $\sigma_{\Delta_{{\rm c}}}$ and $\sigma_{V_{{\rm bc}}}$ are
the standard deviations of $\Delta_{{\rm c}}$ and ${\bf V}_{{\rm bc}}$
projected onto one Cartesian-coordinate axis, respectively. With our
setup, we find that $\sigma_{\Delta_{{\rm c}}}=0.0042$ and $\sigma_{V_{{\rm bc}}}=17.8\,$km/s
at $z_{i}$ (or the root-mean-square of ${\bf V}_{\rm bc}$ is $\sqrt{3}\sigma_{V_{{\rm bc}}}=30.9\,$km/s). At the minimal level of only allowing the variance in
$\Delta_{{\rm c}}$ and $V_{{\rm bc}}$, the average power spectrum
$P_{{\rm m}}(k)$ will then be given by the ensemble average
\begin{eqnarray}
P_{{\rm m}}(k)&=&\int_{0}^{\infty}dV_{{\rm
    bc}}\int_{-\infty}^{\infty}d\Delta_{{\rm c}}\mathcal{P}(k;\,
V_{{\rm bc}})\mathcal{P}(k;\,\Delta_{{\rm c}})\nonumber \\
&&\times P_{{\rm loc,m}}(k;\, V_{{\rm bc}};\,\Delta_{{\rm c}}),
\label{eq:Pk_average}
\end{eqnarray}
where the PDFs and integral arguments are the ones at $z_{i}$. Of
course, a more accurate and straightforward way is to just
ensemble-average 
$P_{{\rm loc,m}}=P_{{\rm loc,m}}\left(k;\, V_{{\rm bc}};\,\Delta_{{\rm c}};\,\Delta_{{\rm b}};\,\Theta_{{\rm c}};\,\Theta_{{\rm b}};\,\Delta_{T};\,\Delta_{T_\gamma}\right)$ over the
patches from a large-box realization,  
because for example $\Delta_{{\rm c}}$ and $\Delta_{{\rm b}}$ are too poorly
correlated at $z_{i}$. 

We numerically integrate Equation (\ref{eq:perturbation_k}) to examine
the evolution of small-scale (high-${\bf k}$) fluctuations at any
overdense (underdense) patch to the linear order, with the help of
Equations (\ref{eq:background_sol}) and (\ref{eq:DeltaT_fitting})
for the evolution of background quantities. In practice, we used the
ODE45 modules of MATLAB\textregistered (2015b, The MathWorks, Inc.,
Natick, Massachusetts, United States) and of GNU Octave, which use
the 4th-order Runge-Kutta method, with the relative tolerance $10^{-4}$
and the absolute tolerance $10^{-2}\delta_{{\bf k}}(a_{i})$. During
the evolution, the number of integration steps is the highest for
$\delta_{T}$, because its amplitude changes from the initial, very
small values around $\delta_{T_{\gamma}}$ to final, much larger values
close to $\delta_{{\rm b}}$. Therefore, taking sub-steps for $\delta_{T_{\gamma}}$
while coarser steps for other $\delta_{{\bf k}}$'s is expected to
boost the computational efficiency, even though we did not yet implement
the method in our computation. The end result is then ensemble-averaged
over varying $\Delta_{{\rm c}}$ and ${\bf V}_{{\rm bc}}$ to obtain
$P_{{\rm m}}(k)$ (Equation \ref{eq:Pk_average}).

\section{Result}
\label{sec:Result}

\subsection{Power spectrum of the matter density}
\label{sub:Pk}

\begin{figure*}
\includegraphics[width=0.325\textwidth]{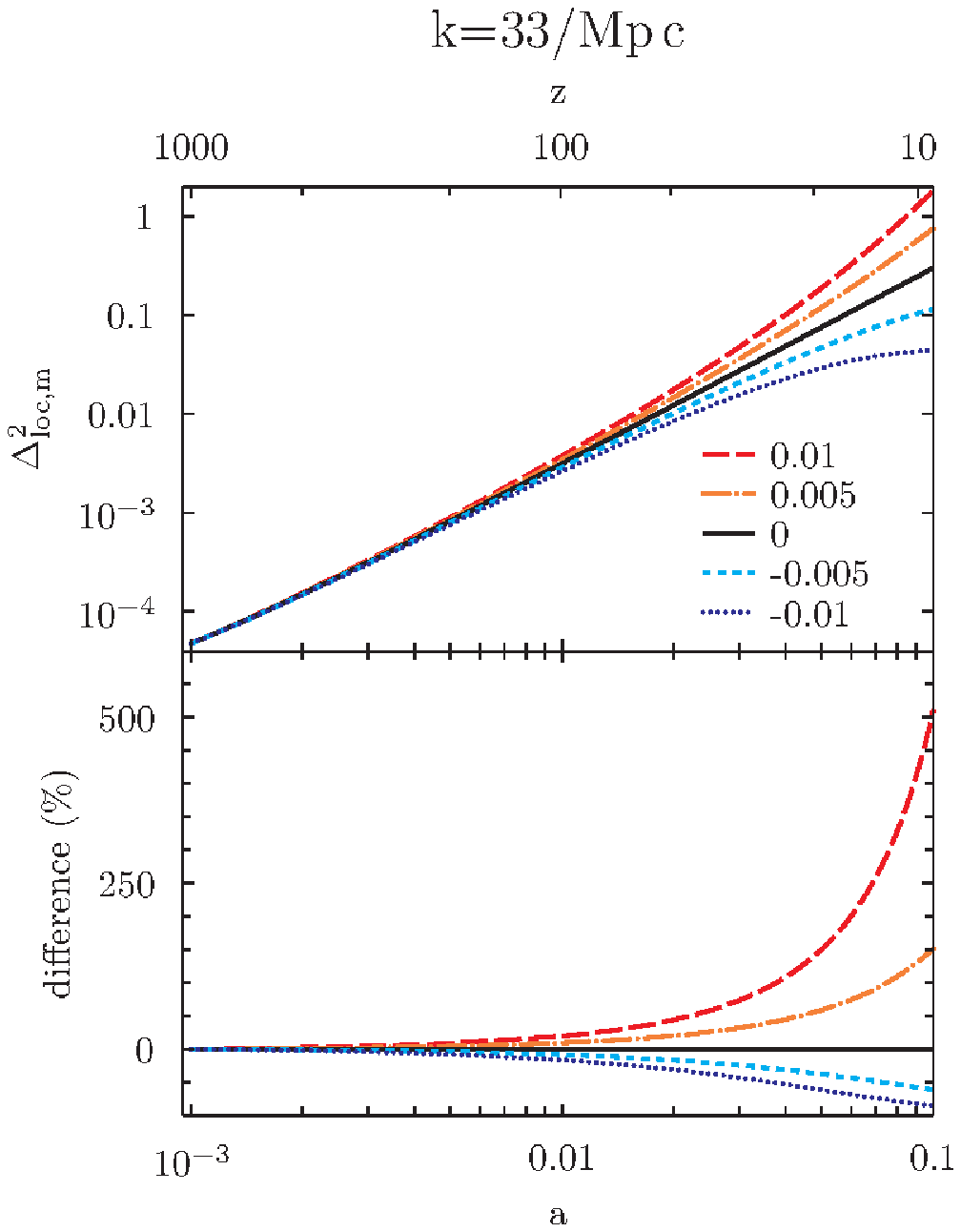}
\includegraphics[width=0.325\textwidth]{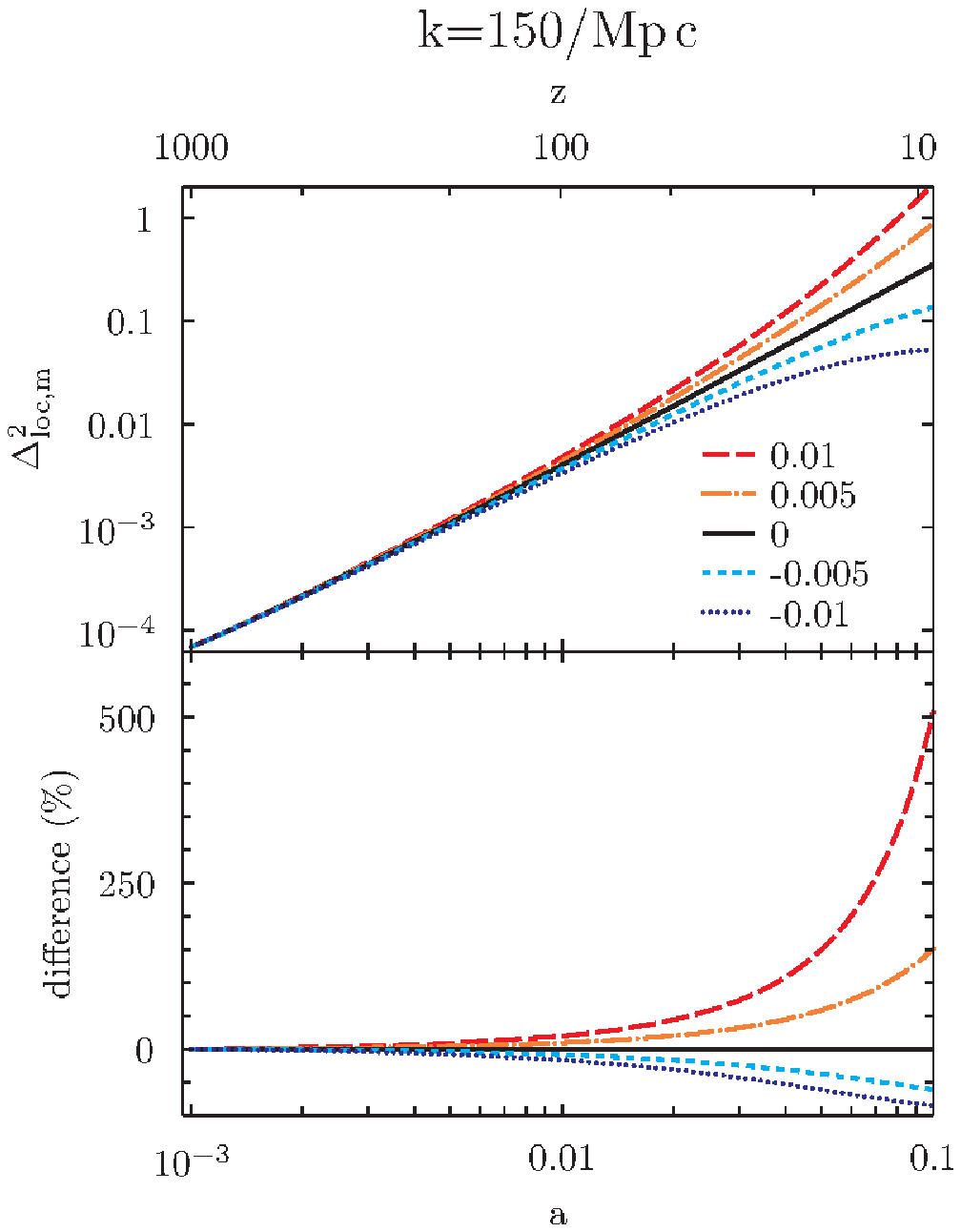}
\includegraphics[width=0.325\textwidth]{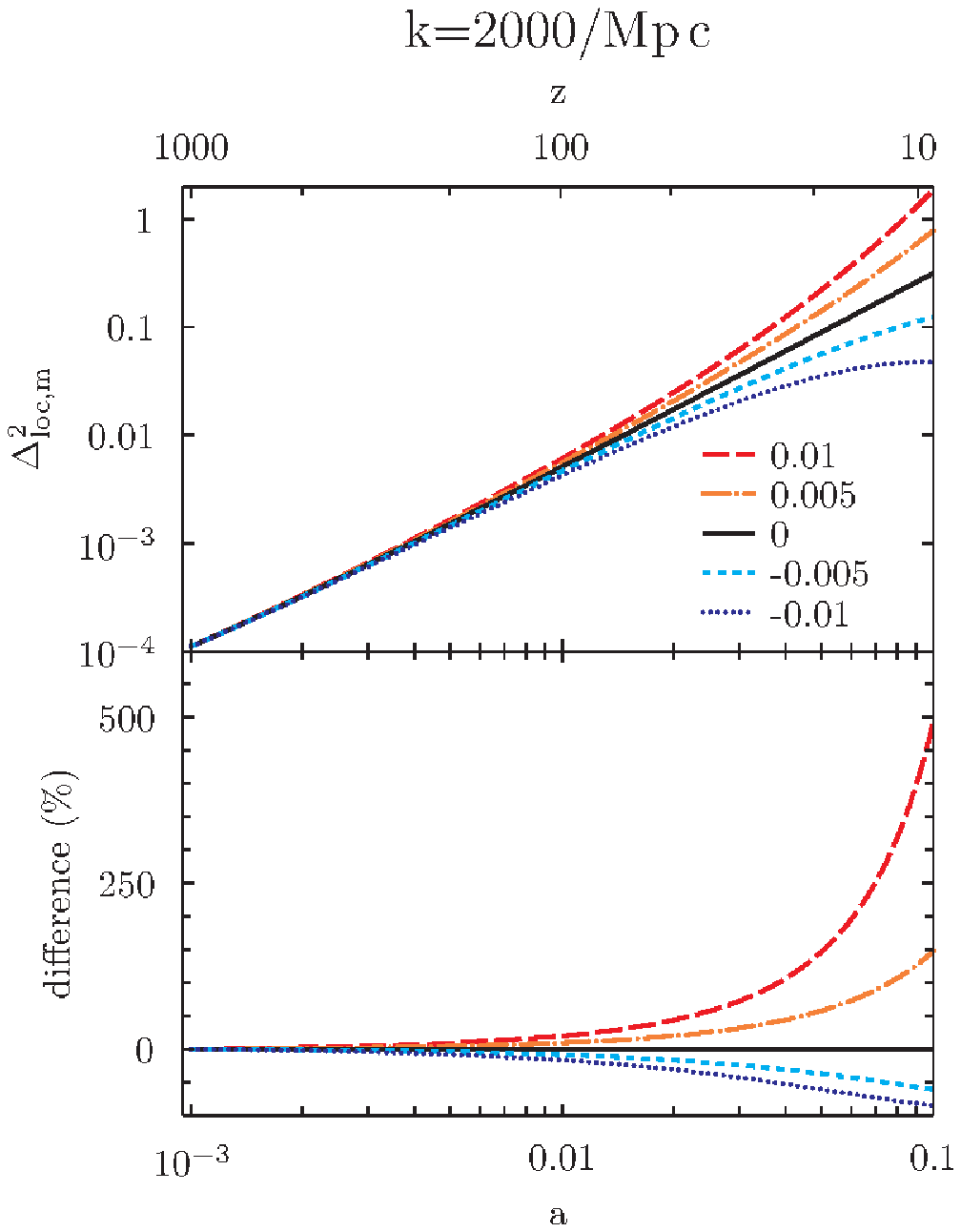}

\caption{Growth of $\Delta_{{\rm loc,m}}^{2}\equiv k^{3}P_{{\rm loc,m}}(k;\, V_{{\rm bc}};\,\Delta_{{\rm c}})/(2\pi^2)$
with wave-numbers $k=$\{33, 150, 2000\}$\,{\rm Mpc}^{-1}$ and $V_{{\rm bc}}(z=1000)=22\,$km/s
in initially overdense (red, long-dashed and orange, dot-dashed),
mean-density (black, solid), and underdense (cyan, short-dashed and
blue, dotted) regions. Initial CDM overdensities are chosen to be $\Delta_{{\rm c}}(a_{i})=$\{-0.01,
-0.005, 0, 0.005, 0.01\}. The mean-density case, or $P_{{\rm loc,m}}(k;\, V_{{\rm bc}};\,\Delta_{{\rm c}}=0)$,
corresponds to $P_{{\rm loc,m}}(k;\, V_{{\rm bc}})$ by TH. Fractional
differences $\left[P_{{\rm loc,m}}(k;\, V_{{\rm bc}};\,\Delta_{{\rm c}})-P_{{\rm loc,m}}(k;\, V_{{\rm bc}};\,\Delta_{{\rm c}}=0)\right]/P_{{\rm loc,m}}(k;\, V_{{\rm bc}};\,\Delta_{{\rm c}}=0)$
in \% are plotted in the bottom sub-panels, with the line-type convention
same as in the top sub-panels. }
\label{fig:Pk_growth}
\end{figure*}

We first examine how the evolution of $P_{{\rm loc,m}}\left(k;\, V_{{\rm bc}};\,\Delta_{{\rm c}}\right)$
depends on the density environment, and compare the result to the
prediction by TH. Figure \ref{fig:Pk_growth} shows the evolution
of $\Delta_{{\rm loc,m}}^{2}\equiv k^{3} P_{{\rm loc,m}}/2\pi^{3}$ of three arbitrarily chosen wave numbers
($k$=\{33,~150,~2000\}~/Mpc) when $V_{{\rm bc}}=22\,{\rm km/s}\,(a/a_{i})^{-1}$
in different density environments ($\Delta_{{\rm c}}(a_{i})$=\{-0.01,~-0.005,~0,~0.005,~0.01\}).
Note again that $\sigma_{\Delta_{{\rm c}}}=0.0042$ at $z=1000$,
and thus these samples correspond to $\pm2.4\sigma_{\Delta_{{\rm c}}}$
and $\pm1.2\sigma_{\Delta_{{\rm c}}}$. First, as expected, the growth
of small-scale fluctuations are biased when $\Delta_{{\rm c}}>0$
and anti-biased when $\Delta_{{\rm c}}<0$, with respect to the mean-density
case (prediction by TH). Secondly, when $\Delta_{{\rm c}}(a_{i})$'s
are equal in amplitude but opposite in sign, the deviations of $P_{{\rm loc,m}}\left(k;\, V_{{\rm bc}};\,\Delta_{{\rm c}}\right)$
from $P_{{\rm loc,m}}\left(k;\, V_{{\rm bc}};\,\Delta_{{\rm c}}=0\right)$
reveal the same trend but only until $z\simeq850$ when $\left|\Delta_{{\rm c}}(a_{i})\right|=0.01$
and $z\simeq730$ when $\left|\Delta_{{\rm c}}(a_{i})\right|=0.005$.
Afterwards, the bias and the anti-bias are not balanced by more than
1\% and such off-balance keeps growing in time. The higher the $\left|\Delta_{{\rm c}}\right(a_{i})|$
is, the earlier this unbalance starts. Thirdly, the fractional deviation
from the mean-density case is almost universal regardless of the value
of $k$, and thus the timing of the unbalance is approximately a function
only of $\left|\Delta_{{\rm c}}(a_{i})\right|$.
Finally, in some high-$\Delta$ patches, our linear analysis based
  on Equation (\ref{eq:perturbation_k}) without quadratic and
  higher-order terms in $\delta$ starts to break down at $z\sim 20$,
  because these modes enter the nonlinear regime ($\Delta_{{\rm
      loc,m}}^{2}(k)\gtrsim 0.1$; see Figure \ref{fig:Pk_growth}) at
  this epoch. A similar breakdown of the
  formalism will occur for perturbations inside very low-$\Delta$ patches as
  well. Therefore, a higher-order scheme than our work is required
  when one is to predict the low-redshift evolution of $\delta$'s. On
  the other hand, if one
  were to generate initial conditions for numerical simulations in the
  linear regime, our formalism would provide the sufficient accuracy.

How large-scale overdensity impacts the evolution of small-scale inhomogeneities
is reflected in the density continuity equation. In overdense background
patches, $\Delta_{j}$ grows in time and $\Theta_{j}<0$. Then, in
Equation (\ref{eq:perturbation_k}), $-\Delta_{j}\theta_{j}$ and
$-\Theta_{j}\delta_{j}$ work as sources terms in addition to $-\theta_{j}$
to the growth of $\delta_{j}$. This will boost the growth rate of
$\delta_{j}$ of both overdense ($\delta_{j}>0$, $\theta_{j}<0$)
and underdense ($\delta_{j}<0$, $\theta_{j}>0$) modes. In contrast,
in underdense background patches, these terms suppress the growth
rate of $\delta_{j}$  of both overdense and underdense modes. The
distribution of $\Delta_{j}(a_{i})$ is Gaussian, and therefore for
each ``bias'' case with $\Delta_{j}>0$ there exists an ``anti-bias''
case with $\Delta_{j}<0$. Nevertheless, the unbalance described above
is expected to boost the average power spectrum from that by TH.

\begin{figure*}
\includegraphics[width=0.45\textwidth]{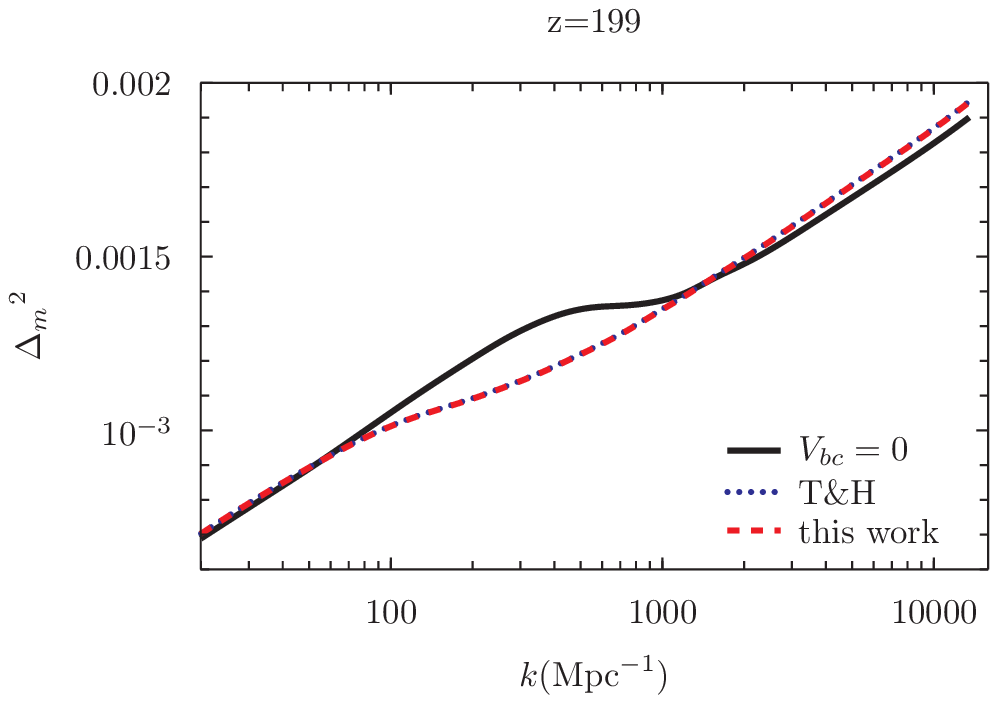}
\includegraphics[width=0.45\textwidth]{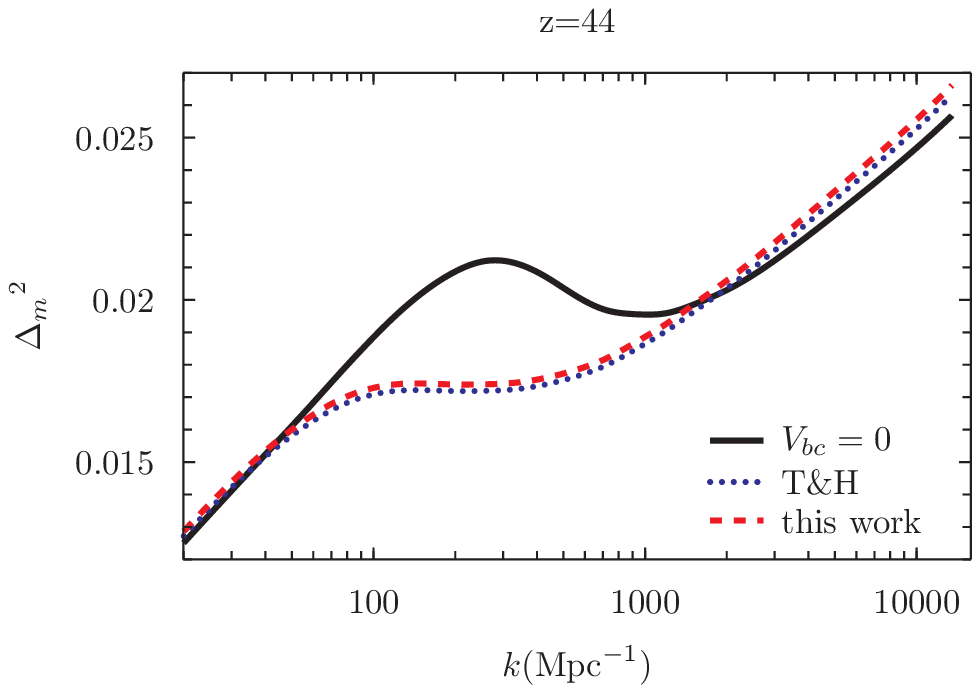}

\includegraphics[width=0.45\textwidth]{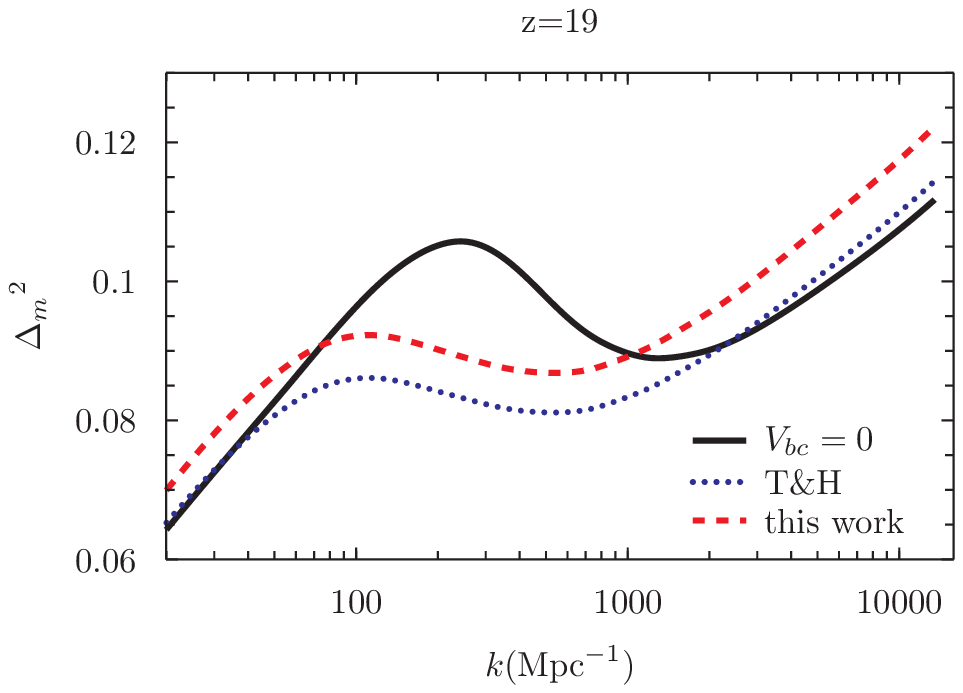}
\includegraphics[width=0.45\textwidth]{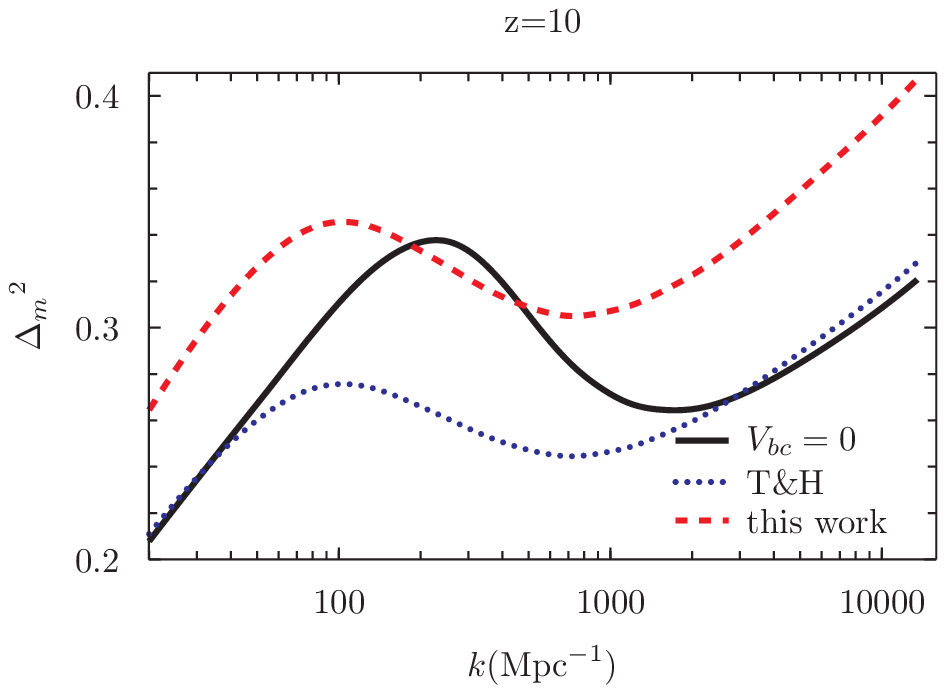}

\caption{Mean power spectrum $P_{m}(k)$ of the matter-density fluctuation,
expressed in terms of the ${\bf k}$-space variance $\Delta_{m}(k)$
(defined in the text). Comparison is made for the case without the
relative-velocity effect (black, solid), the case investigated by
TH (blue, dotted) and the case investigated in this work (red, short-dashed).}
\label{fig:Pk_compare}
\end{figure*}

The overall effect of including the Gaussian distribution of $\Delta_{{\rm c}}$
is thus to mitigate the negative impact by the relative velocity,
predicted by TH, to some extent. In addition, the universality of
the unbalance in ${\bf k}$ boosts $P_{{\rm m}}(k)$ even in the ${\bf k}$
range ($10\lesssim{\bf k}\lesssim100/{\rm Mpc}$ and ${\bf k}\gtrsim1000/{\rm Mpc}$)
where the power spectrum is almost unaffected by non-zero $V_{{\rm bc}}$
(Figure \ref{fig:Pk_compare} shown in terms of the ${\bf k}$-space
matter-density variance
$\Delta_{m}^{2}\equiv k^{3}P_{{\rm m}}(k)/2\pi^{2}$). Note that the
result for ${\bf k}\lesssim10/{\rm Mpc}$ cannot be trusted, because
our perturbation theory is based on the condition that large-scale
modes ($0.01\lesssim{\bf K}\,{\rm Mpc}\lesssim1$) are well separated
from small-scale modes in scale. Discrepancy of $P_{{\rm m}}(k)$
including non-zero $\Delta$'s from the prediction by TH is negligible
at $z\gtrsim45$, but later the discrepancy grows in time. Of course,
individual patches may experience discrepancy in $P_{{\rm loc,m}}(k;\,
V_{{\rm bc}};\,\Delta_{\rm c})$
much earlier than this epoch (Figure \ref{fig:Pk_growth}). 

\subsection{Halo abundance}
\label{sub:mass_function}

Understanding the abundance and the spatial distribution of cosmological
halos is crucial in modern astrophysics and cosmology. In this section,
we examine the halo abundance both in the local and the global sense,
just as we did for the matter power spectrum.

Let us first revisit the calculation by TH. They adopted the extended
Press-Schechter formalism and calculated the local halo abundance,
in terms of the conditional mass function, using the peak-background
split scheme. A patch with $\Delta$ (let us use this notation for
the matter overdensity, to avoid confusion with the ${\bf k}$-space
matter-density variance $\Delta_{m}^{2}$) and $V_{{\rm bc}}$ will
have the number of  halos per unit Eulerian comoving
volume per $M$ given by

\begin{eqnarray}
\frac{dn}{dM}(M|\Delta,\, V_{{\rm
    bc}})&=&\sqrt{\frac{2}{\pi}}\frac{\bar{\rho}_{m}}{M}\frac{\delta_{{\rm
      crit}}-\Delta}{\sigma^{2}}\left|\frac{d\sigma}{dM}\right|\left(1+\Delta\right)
\nonumber \\
&&\times \exp\left[-\frac{\left(\delta_{{\rm
        crit}}-\Delta\right)^{2}}{2\sigma^{2}}\right],
\label{eq:cond_mass_ftn}
\end{eqnarray}
where $\delta_{{\rm crit}}$ is the critical overdensity of spherical
collapse, and $\sigma^{2}$ is the variance of density field smoothed
with the window function $W_{M}$ corresponding to mass $M$,

\begin{equation}
\sigma^{2}(M,\, V_{{\rm bc}})=\int\Delta_{m}^{2}(k,\, V_{{\rm bc}})W_{M}^{2}\, d\ln k.\label{eq:sigma2_TH}
\end{equation}
One should note that $\sigma^{2}(M,\, V_{{\rm bc}})$ should be that
of high $k$ modes only, or more accurately a reduced value $\sigma^{2}(M,\, V_{{\rm bc}})-\sigma_{{\rm patch}}^{2}$
where $\sigma_{{\rm patch}}^{2}$ is the variance of density field
smoothed with the window function corresponding to the mass of the
patch%
\footnote{It is not clear whether TH used this reduced variance. In addition,
a factor of 2 should be multiplied to Equation (18) of TH.%
} (e.g. \citealt{Bond1991}; \citealt{Mo1996}; \citealt{Ahn2015}).
One can instead put a lower bound $k_{{\rm min}}$ ($=\left[6\pi^{2}\bar{\rho}_{m}/M_{{\rm patch}}\right]^{1/3}$)
in the integral of Equation (\ref{eq:sigma2_TH}), which would be
identical to the reduced variance if a sharp $k$-space window function
is used. The global mass function $(dn/dM)_{{\rm g}}$ is simply an
average of the local mass function, $dn(M|\Delta,\, V_{{\rm bc}})/dM$,
over the ensemble of patches. When $dn(M|\Delta,\, V_{{\rm bc}})/dM$
is averaged only over $V_{{\rm bc}}$ for a given $\Delta$, which
is equivalent to visiting only those patches with the same $\Delta$
and taking the average, it leads to the conditional mass function
$(dn/dM)_{\Delta}\equiv dn(M|\Delta)/dM$.

This calculation should be modified, because $\sigma^{2}$ depends
also on $\Delta$ through the dependence of $P_{{\rm loc,m}}$ on
$\Delta$:

\begin{equation}
\sigma^{2}(M,\, V_{{\rm bc}},\,\Delta)=\int_{k_{{\rm min}}}^{k_{{\rm max}}}\Delta_{m}^{2}(k,\, V_{{\rm bc}},\,\Delta)\, d\ln k,\label{eq:sigma2_Ahn}
\end{equation}
which will enter Equation (\ref{eq:cond_mass_ftn}). Here we used
the sharp ${\bf k}$-space filter, and thus $k_{{\rm max}}=\left[6\pi^{2}\bar{\rho}_{m}/M\right]^{1/3}$.
An overdense patch will then have a boost in $dn(M|\Delta,\, V_{{\rm bc}})/dM$
from the value by TH because $P_{m}(k,\, V_{{\rm bc}},\,\Delta>0)>P_{m}(M,\, V_{{\rm bc}},\,\Delta=0)$
and thus $\sigma^{2}(M,\, V_{{\rm bc}},\,\Delta>0)>\sigma^{2}(M,\, V_{{\rm bc}},\,\Delta=0)$,
and vice versa (a decrease from the value by TH) for an underdense
patch. Obviously, both $(dn/dM)_{\Delta}$ and $(dn/dM)_{{\rm g}}$
will also be affected.

It is important to compare our findings to the usual peak-background
split scheme and the one by TH. In the ``standard'' scheme, if the
density field is purely Gaussian, all the wave modes are assumed mutually
independent in the linear regime. Therefore, the local, high-${\bf k}$
modes have a universal%
\footnote{Rigorously speaking, it is not perfectly universal because the lower
bound changes slightly in $\Delta$ as $M_{{\rm patch}}=\bar{\rho}_{m}(1+\Delta$). %
} variance $\sigma_{{\bf k}}^{2}$ whether or not they are placed inside
a patch with non-zero $\Delta$. The way how the halo formation is
biased in an overdense region is simply through the shift in the density
($+\Delta$). TH then realized the fact that the variance is not universal
but should depend on $V_{{\rm bc}}$. Because non-zero $V_{{\rm bc}}$
tends to suppress $\sigma_{{\bf k}}^{2}$, the odds to cross $\delta_{{\rm crit}}$
decrease relative to the standard picture. We find that there is another
dependency of the variance, which is $\Delta$. In other words, we
find that there are two biasing effects in an overdense region compared
to a mean-density region: getting closer to $\delta_{{\rm crit}}$
because of the shift in the density ($+\Delta$, also in the standard
scheme), and having a larger degree of fluctuation in $\delta_{{\bf k}}$
due to the mode-mode coupling (e.g. source terms $-\Delta_{{\rm c}}\theta_{{\rm c}}$
and $-\Theta_{{\rm c}}\delta_{{\rm c}}$ in $\partial\delta_{{\rm c}}/\partial t$
in Equation \ref{eq:perturbation_k}, which is a new finding). Therefore,
by not fully implementing the effect of non-zero $\Delta$, TH in effect
underestimates and overestimates the halo mass functions in overdense
and underdense regions, respectively. 

We note that this additional bias effect  should be 
  present even in the standard picture with $V_{{\rm bc}}=0$, because
  this is due to the natural coupling between the large-scale and
  small-scale 
  density perturbations. In this case, however, we are not sure about
  the quantitative validity of the extended Press-Schechter formalism on the
  conditional mass function (Equation \ref{eq:cond_mass_ftn}), which
  is based on the linear theory guaranteeing Gaussianity at any filtering
  scales {\em without} the mode-mode
  couplings. Qualitatively, we believe that the boost of local
  $\delta_{{\bf k}}$ and $\sigma_{{\bf k}}^{2}$ under $\Delta$ should
  boost the conditional mass function to the level estimated by
  Equations (\ref{eq:cond_mass_ftn}) and (\ref{eq:sigma2_Ahn}) as
  described above anyways. We defer a 
  further investigation of this issue, which can be clarified with numerical
  simulations of the halo formation under different $\Delta$'s.

\begin{figure*}
\gridline{
\fig{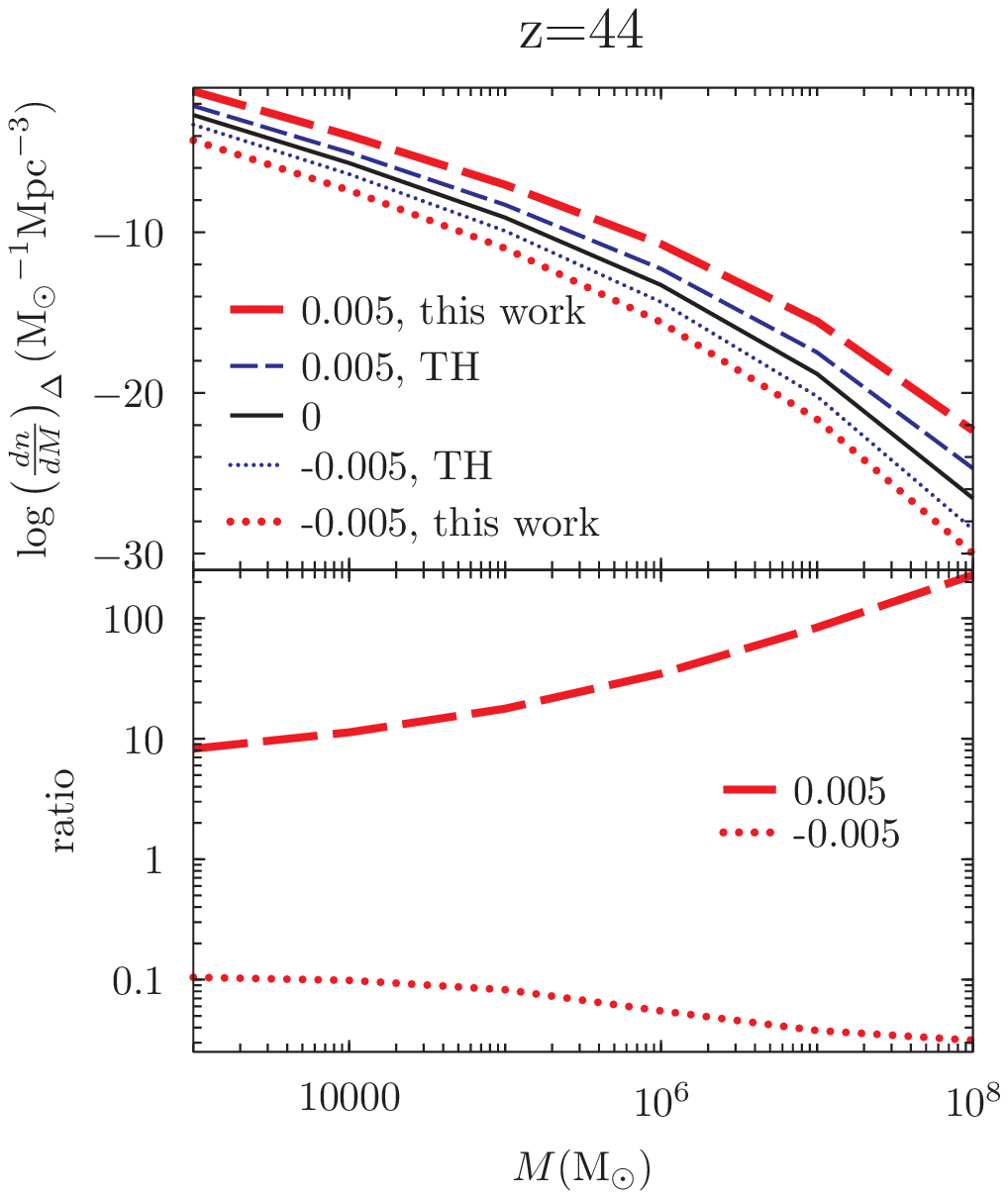}{0.3\textwidth}{}
\fig{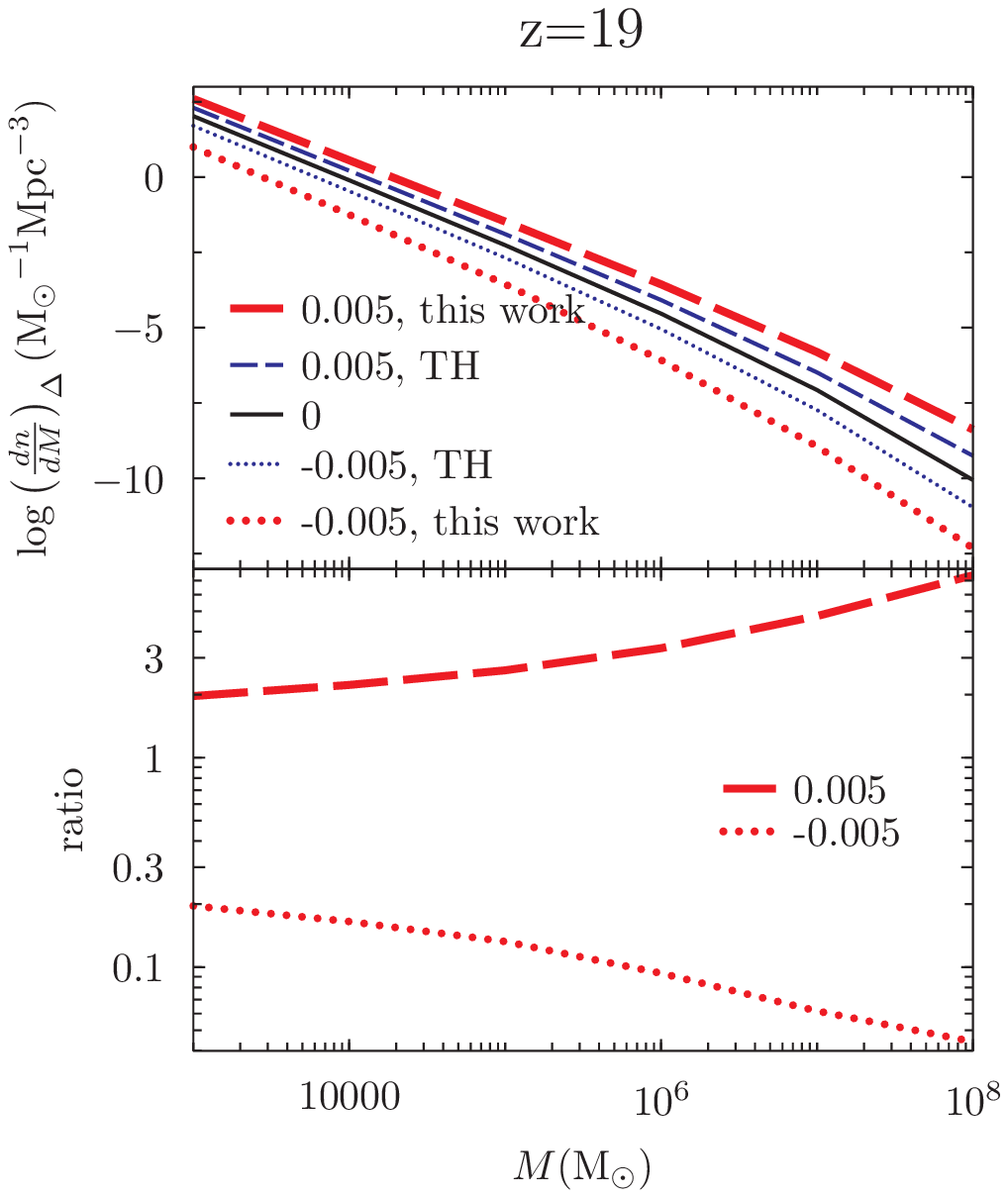}{0.3\textwidth}{}
}

\gridline{
\fig{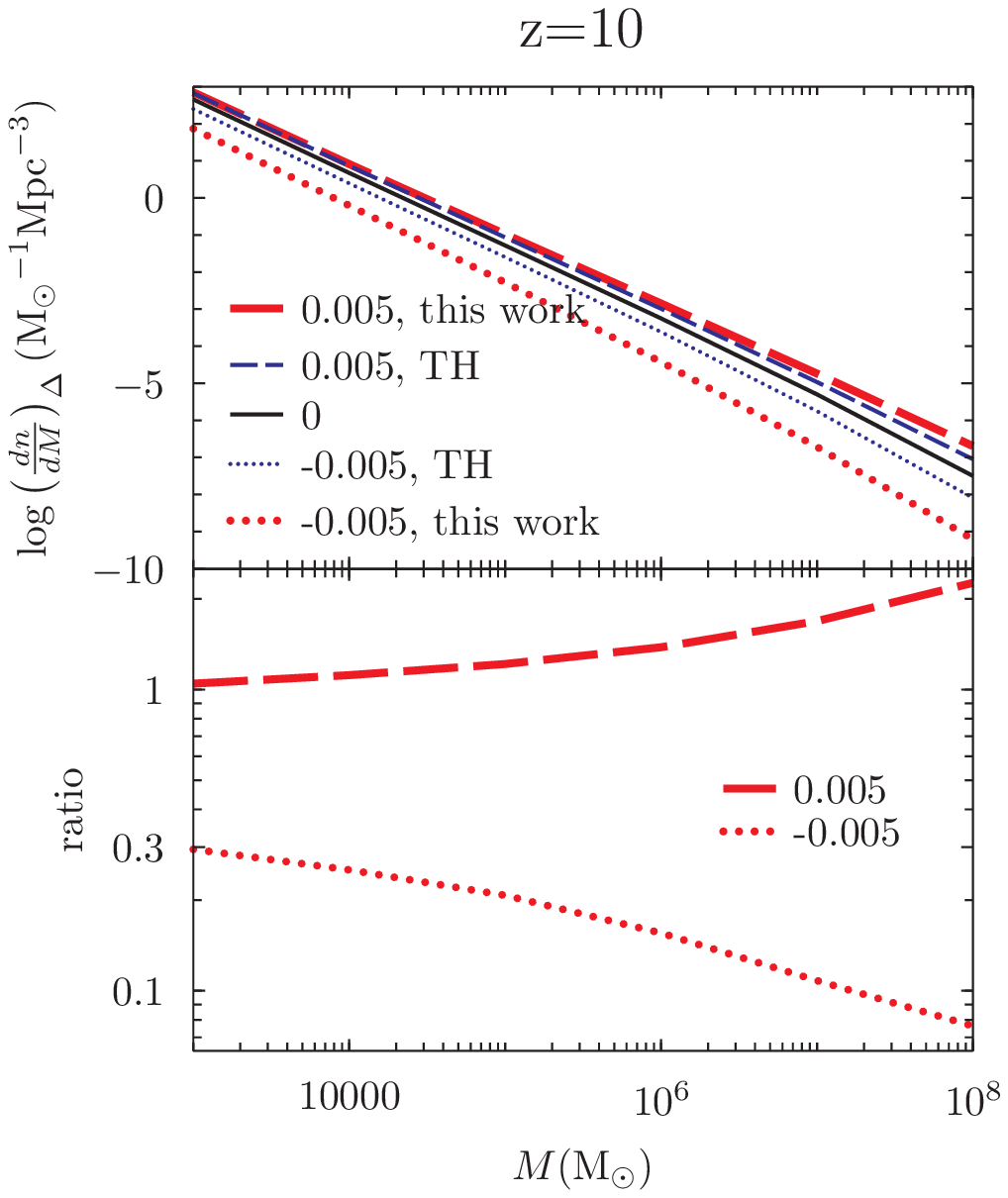}{0.3\textwidth}{}
\fig{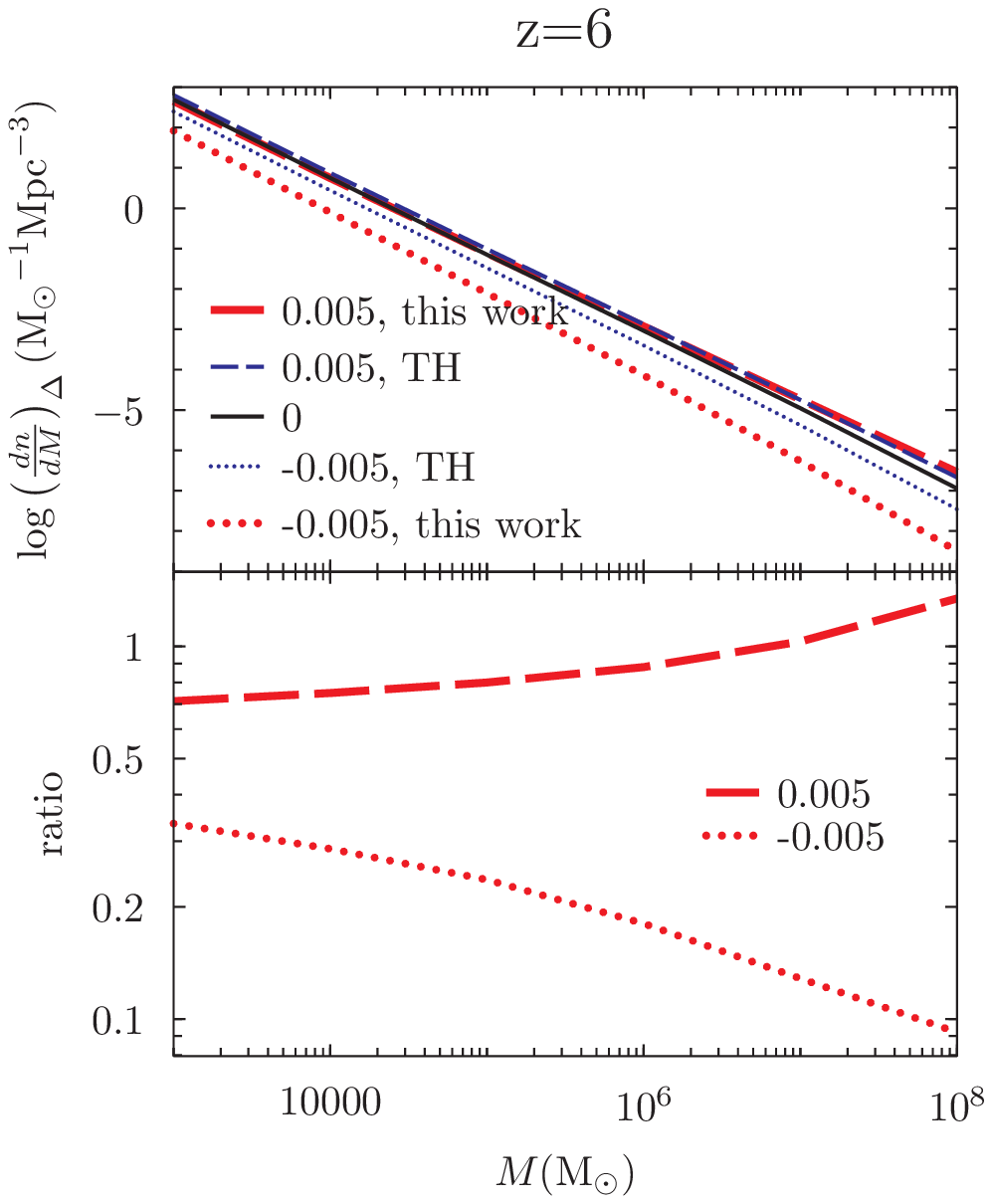}{0.3\textwidth}{}
}

\caption{Conditional halo mass functions, at varying redshifts (in title),
for patches with different initial overdensities: $\Delta_{{\rm c}}(a_{i})$
= -0.005 (dotted), 0 (solid) and 0.005 (dashed). Thick (red) curves
represent our predictions, and thin curves (blue) represent predictions
by TH. There is no distinction between the predictions when $\Delta_{{\rm c}}(a_{i})=0$,
because then $V_{{\rm bc}}$ is the only variable for the ensemble
in both cases. In each bottom sub-panel, we show the ratio
of our prediction to the prediction by TH, or $(dn/dM)_{\Delta}/(dn/dM)_{\Delta,\,{\rm TH}}$.}
\label{fig:cond_mass_ftn}
\end{figure*}

Discrepancy between the conditional mass functions by this work and
by TH is significant if we focus on individual patches. Figure \ref{fig:cond_mass_ftn}
illustrates how our prediction differs from that by TH. For example,
at $z\sim 44-19$, under $\Delta_{{\rm c}}(a_{i})=0.005$
we predict {[}100 - 2000{]} \% boost in $(dn/dM)_{\Delta}$ compared
to the values by TH (let us denote them by $(dn/dM)_{\Delta,\,{\rm TH}}$).
For $\Delta_{{\rm c}}(a_{i})=-0.005$, we predict 90 \% or more
decrease in $(dn/dM)_{\Delta}$ compared to the values by TH. This
is the obvious result of the mode-mode coupling of $\Delta$ and
$\delta$ described above. 
It is also noteworthy that the discrepancy is the largest
for the rarest halos: first, at any redshift, the discrepancy increases
as the halo mass increases and secondly, for any given halo mass,
the discrepancy decreases in time.

\begin{figure*}
\includegraphics[width=0.32\textwidth]{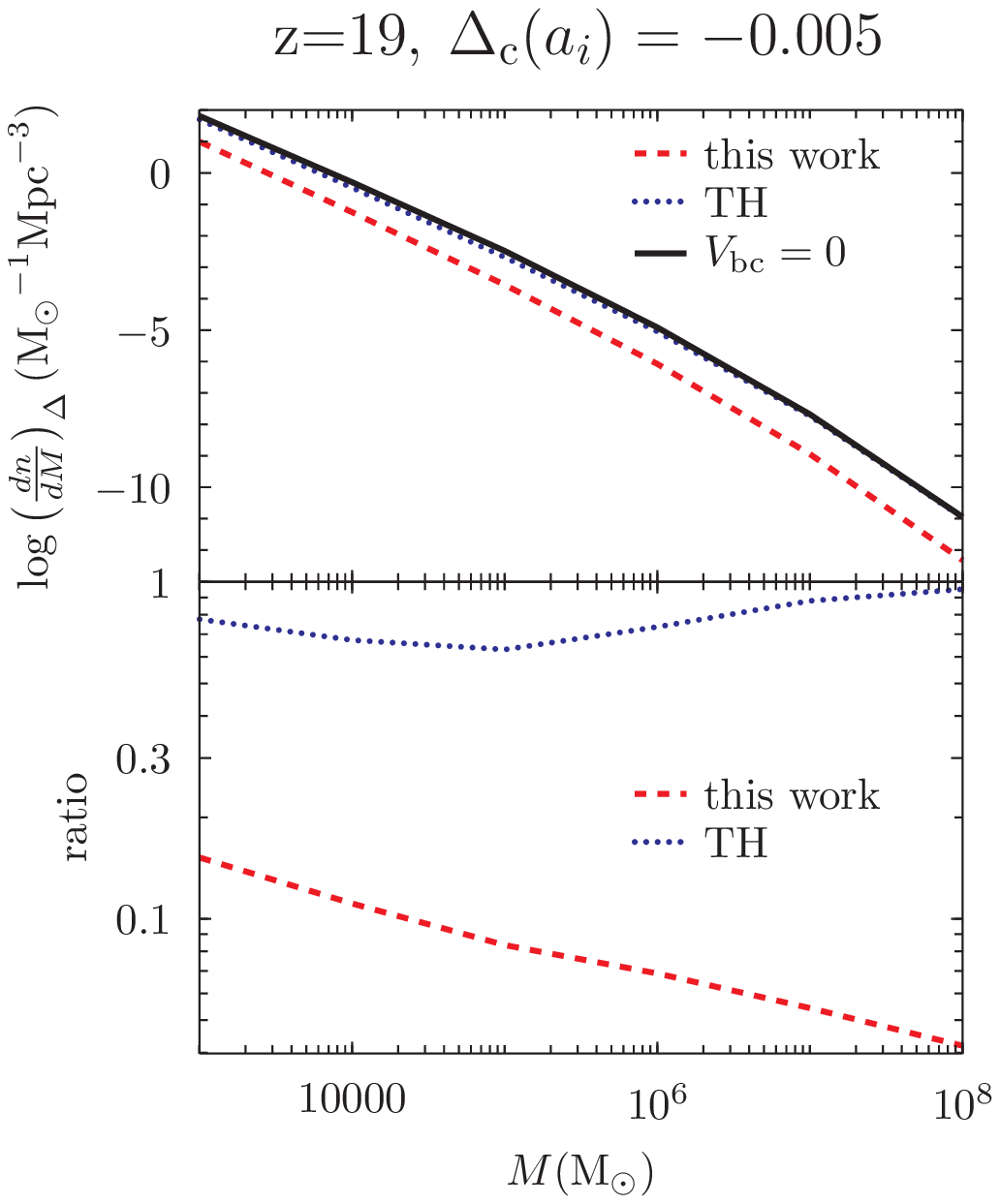}
\includegraphics[width=0.32\textwidth]{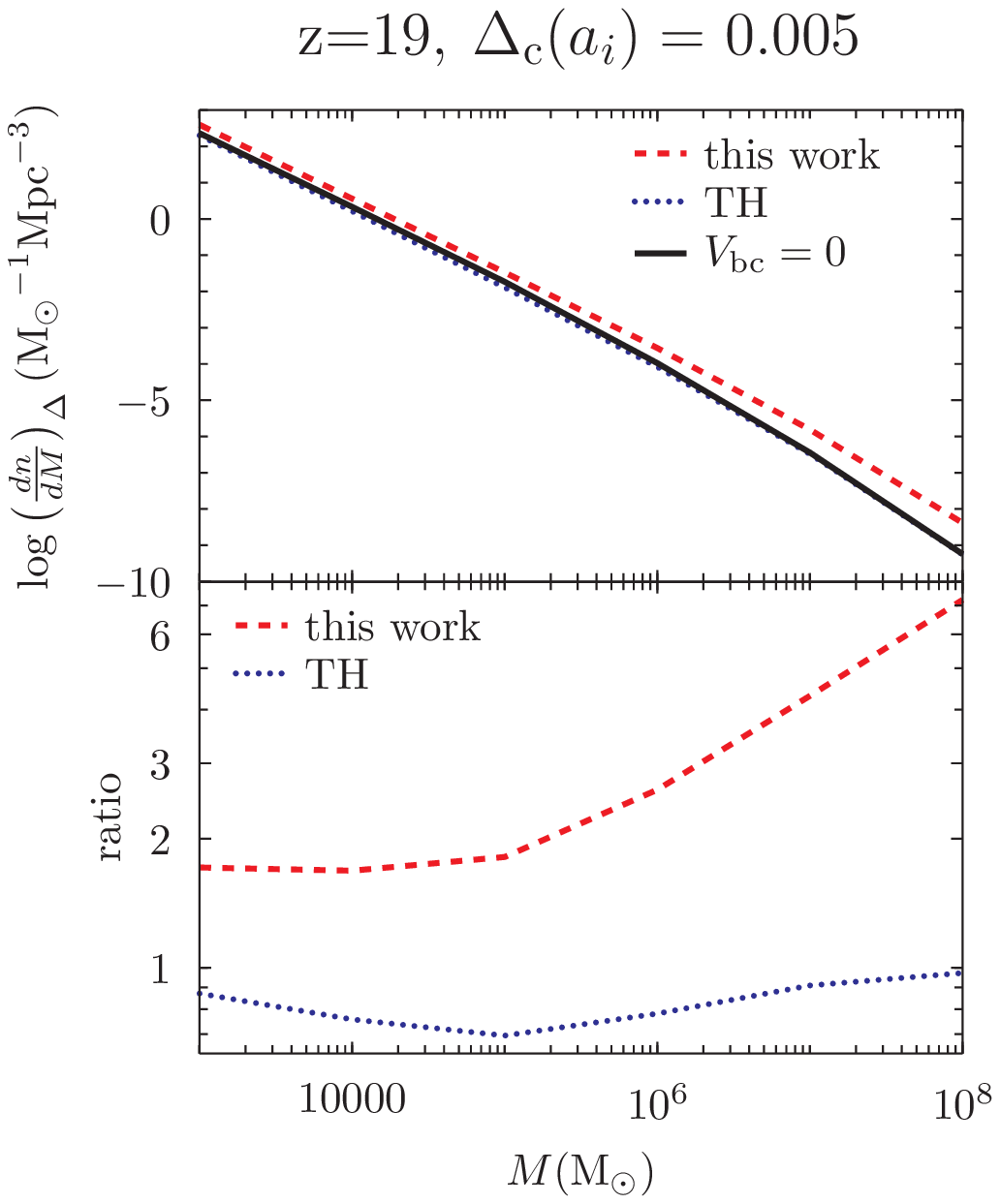}
\includegraphics[width=0.32\textwidth]{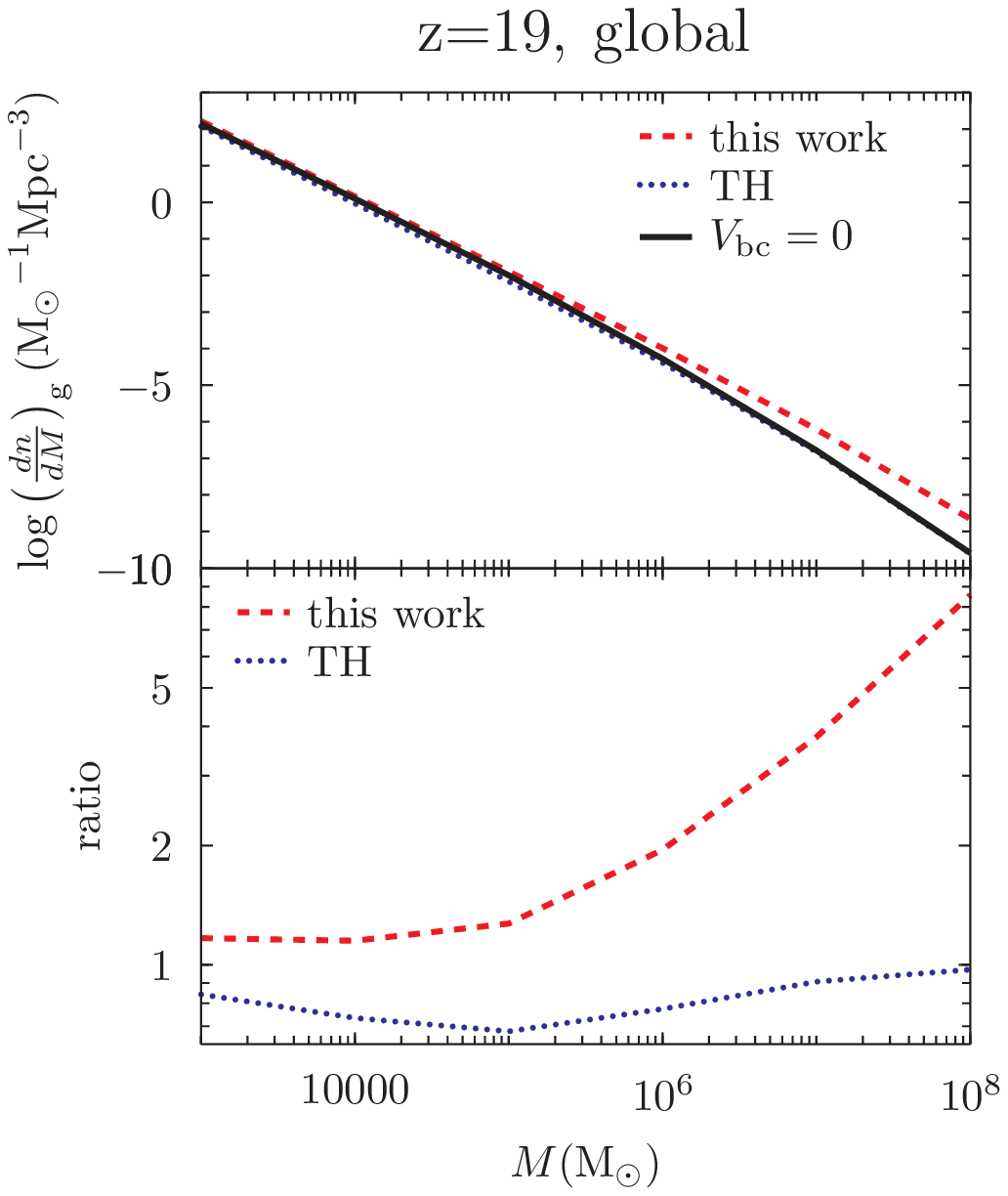}

\caption{Conditional mass functions at $z=19$ with $\Delta_{{\rm c}}(a_{i})$
= -0.005 (left) and 0.005 (middle). Comparison is made among this
work (dashed; red), TH (dotted; blue) and the standard picture with
$V_{{\rm bc}}=0$ (solid; black). Global mass functions are plotted
on the right panel. In each bottom sub-panel, we show the ratio of
mass functions to that of the standard picture, 
or $(dn/dM)/(dn/dM)_{{\rm V_{{\rm bc}}=0}}$.}
\label{fig:global_mass_ftn}
\end{figure*}

The discrepancy among the conditional mass functions by this work,
by TH and by the standard picture also influences $(dn/dM)_{{\rm g}}$.
Let us just take the example of $z=19$
(Figure \ref{fig:global_mass_ftn}).
Compared to the standard prediction ($V_{\rm bc}=0$), conditional mass functions by
TH stay lower regardless of $\Delta$. This is due to the suppression
of structure formation by the relative velocity. In contrast, $(dn/dM)_{\Delta}$
in this work is either higher or lower than the standard prediction
($\equiv(dn/dM)_{{\rm V_{{\rm bc}}=0}}$) depending on the halo mass
and $\Delta$. At $z=19$, for $\Delta_{{\rm c}}(a_{i})=-0.005$
$(dn/dM)_{\Delta}<(dn/dM)_{{\rm V_{{\rm bc}}=0}}$ and for $\Delta_{{\rm c}}(a_{i})=0.005$
$(dn/dM)_{\Delta}>(dn/dM)_{{\rm V_{{\rm bc}}=0}}$ for any halo mass.
The tendency for $(dn/dM)_{\Delta}$ to overshoot $(dn/dM)_{{\rm V_{{\rm bc}}=0}}$
when $\Delta>0$ is not generic, because the influence of the positive
$\Delta$'s on the structure formation appears only late in its evolution
(see Figs. \ref{fig:Pk_growth} and \ref{fig:Pk_compare}). At any
rate, the contrast in $(dn/dM)_{\Delta}$ among overdense and underdense
regions is increased compared to TH and the standard picture, and
the net effect e.g. at $z=19$ is to boost $(dn/dM)_{{\rm g}}$
from $(dn/dM)_{{\rm g},\, V_{{\rm bc}}=0}$. At much higher redshifts, our
$(dn/dM)_{{\rm g}}$ is almost indistinguishable from that by TH,
which undershoots $(dn/dM)_{{\rm g},\, V_{{\rm bc}}=0}$.

We note that $(dn/dM)_{\Delta}$ and $(dn/dM)_{{\rm g}}$ have the
usual problem of not correctly predicting the actual mass function, if
one sticks to the original extended Press-Schechter
formalism. Minihalos at high redshifts are usually underestimated by
the extended Press-Schechter formalism.
The usual peak-background split method suffers
from large discrepancies between its prediction and the N-body simulation
results for rare halos in general. In this case, a hybrid method to
connect the peak-background-split halo bias parameter to the
better-fitting mean mass function  
types (e.g. \citealt{2004ApJ...609..474B}; \citealt{Ahn2015}) is much
more appropriate. We will apply this method in the future for a
better estimation of the conditional mass function and the ${\bf
  k}$-space halo bias parameter (Section \ref{sub:Halo-bias}).

\subsection{Halo bias and stochasticity}
\label{sub:Halo-bias}

The conditional mass function we examined in Section \ref{sub:mass_function}
is an indicator of how halo formation is biased toward overdense regions.
The halo bias can be viewed also in the the ${\bf k}$-space. This is
a crucial parameter in cosmology when trying to probe the fluctuation
of the matter density from surveys of galaxies through, for example,
the power spectrum analysis.

\begin{figure}
\includegraphics[width=0.4\textwidth]{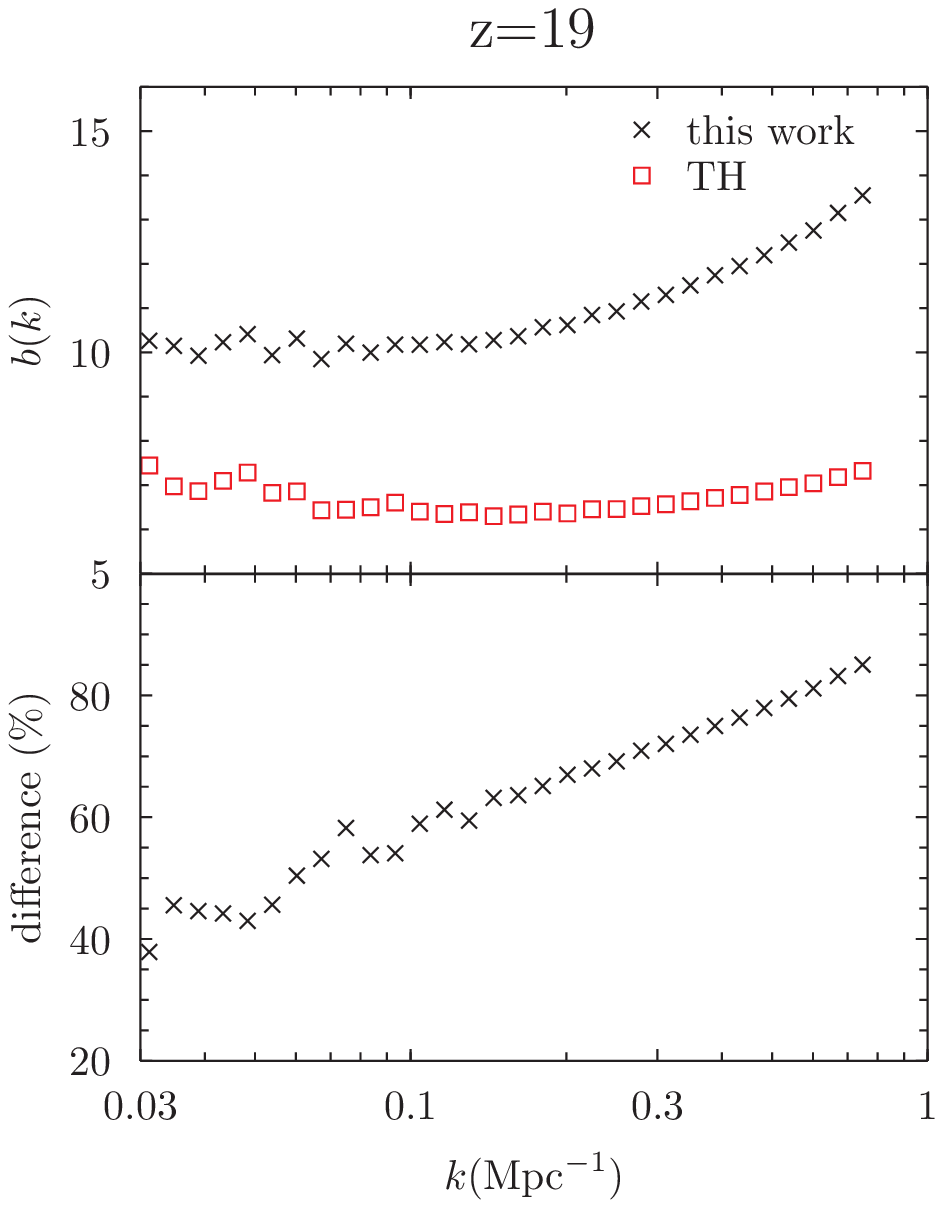}

\caption{Halo bias parameter for halos with $M=10^{6}\, M_{\odot}$ at $z=19$.
Comparison is made between this work (black, x) and TH (red, square)
in the top sub-panel. In the bottom sub-panel, we show the fractional
difference $\left[b(k)-b(k)_{{\rm TH}}\right]/b(k)_{{\rm TH}}$.}
\label{fig:bias}
\end{figure}

The halo bias parameter in ${\bf k}$-space is defined as

\begin{equation}
b(k)=\left(\frac{P_{h}(k)}{P_{m}(k)}\right)^{1/2},\label{eq:bias}
\end{equation}
where $P_{h}(k)$ is the power spectrum of halo over-abundance
\begin{equation}
\delta_{n}(M,\,{\bf x})=\frac{\left(\frac{dn}{dM}\right)_{\Delta}(M,\,{\bf x})-\left(\frac{dn}{dM}\right)_{g}(M)}{\left(\frac{dn}{dM}\right)_{g}(M)}.\label{eq:over-abundance}
\end{equation}
$\delta_{n}(M,\,{\bf x})$ can then be Fourier-transformed in order
to calculate $P_{h}(k)$. We expect $b(k)$ to be larger than that
predicted by TH, because the ``contrast'' in $(dn/dM)_{\Delta}$
between overdense and underdense regions has increased from that by
TH (Section \ref{sub:mass_function}; Figure \ref{fig:cond_mass_ftn}).
This is clearly seen in Figure \ref{fig:bias}, where we show $b(k)$'s
computed in this work and by TH. At $z=19$ and for $M=10^{6}\, M_{\odot}$,
we find that $b(k)$ is about $[1.5-2]$ times as large as the one by
TH. This 
is again caused by the mode-mode coupling.

As was pointed out by TH, $b(k)$ oscillates in $k$ due to the baryonic
acoustic oscillations (BAO)%
\footnote{Our box size is barely larger than the BAO scale, and is thus too
small to accurately estimate $b(k)$ at low $k$'s. We will increase
the size of the box in future work for a better estimation.%
}, which is tied to the modulation of the streaming velocity, the
density fluctuation, and the baryon fraction in the
existence of the compensated mode \citep{Barkana2011}.
This makes it difficult to deduce $P_{m}(k)$ from $P_{h}(k)$,
not to mention from the galaxy surveys where galaxy formation mechanism,
strongly influenced by baryonic physics, is another nuisance parameter
(but see \citealt{Slepian2015} for how to separate out the streaming
velocity effect).
For the halo mass range treated in this paper, $b(k)$ should modulate
the distribution of the first stars which grow predominantly inside
minihalos. As was also noted by TH, the difference in $b(k)$ should
also influence the formation of much larger-mass halos, which are
used for galaxy surveys. It is also possible that the nonlinear effect
described by \citet{O'Leary2012}, or the heating of the intergalactic
medium (IGM) due to the velocity difference between CDM and baryons,
is modulated in space depending on the overdensity. Then the power
spectrum in the 21-cm background, which may be dominated by the velocity
fields if the heating is efficient (\citealt{McQuinn2012}), is likely
to be boosted. Because such a power spectrum shows a very clear BAO
feature and the 21-cm observation usually suffers from the low sensitivity,
the signal boosted even more from the prediction by \citet{McQuinn2012}
will be a very promising target for the high-redshift 21-cm cosmology.

Let us briefly discuss the halo stochasticity. The halo stochasticity
is defined as

\begin{equation}
\chi=\frac{P_{hm}^{2}(k)}{P_{h}(k)\, P_{m}(k)},\label{eq:stochasticity}
\end{equation}
where $P_{hm}(k)$ is the cross power spectrum between the halo density
and the matter density. Both in TH and in this work, stochasticity is caused
by the fluctuation in $(dn/dM)_{\Delta}$ because patches with the
same $\Delta$ can have different $V_{{\rm bc}}$'s which affect $(dn/dM)_{\Delta}$.
One should note that the fluctuation should be caused also by the
sampling variance. The conditional mass functions usually show super-Poissonian
distributions in $(dn/dM)_{\Delta}$ even in the standard picture
with $V_{{\rm bc}}=0$ (e.g. \citealt{Saslaw1984}; \citealt{Sheth1995};
\citealt{Neyrinck2014}; \citealt{Ahn2015}), and obviously this should
cause the stochasticity in addition to that by the varying $V_{{\rm bc}}$.
Because we use the ``mean'' conditional mass function just as TH
did, in both works $\chi$ does not reflect the sampling variance.
Including this effect requires the calculation of the sub-cell correlation
function (\citealt{Ahn2015}), which we delay to future work.

\section{Discussion}
\label{sec:Discussion}

We investigated the impact of the relative velocity (streaming velocity)
between CDM and baryons on the small-scale structure formation. TH
first studied this effect by adopting a trivial solution to the large-scale
velocity and density fields. Because velocity fields are correlated
with density fields, however, such a trivial solution cannot accurately
describe the physics in regions with non-zero overdensity. We thus
improved on the work by TH by implementing a non-trivial solution
to the large-scale velocity and density fields, and we find that this
causes a new type of coupling between large-scale and small-scale
modes. This
results in boosting the small-scale structure formation in overdense
regions and suppressing that in underdense regions, aside from the
suppression originating from the streaming velocity. The net effect
on the structure formation is to boost the overall fluctuation, in
terms of $P_{m}(k)$, and thus the ``negative'' effect by TH is
mitigated to some extent. Depending on the wave mode (${\bf k}$) and
the observing redshift, $P_{m}(k)$ can even be larger than that in
the standard picture with $V_{{\rm bc}}=0$. The conditional halo
mass function and the halo bias are also affected in similar ways.

The results of this work show that the formation and evolution of
small-scale structures depend strongly on not only the streaming velocity
but also the density environment. The most important aspect of our
work is that in contrast to TH, who predict that regardless of the
underlying density the local matter power spectrum of small-scale
structures will be identical as long as $V_{{\rm bc}}$ is the same,
the underlying large-scale ($\sim$ a few Mpc) overdensity
is another key parameter in addition to $V_{{\rm bc}}$. This then
requires re-examining previous work based on the formalism by TH.
We already showed that $P_{{\rm loc,m}}(k)$, $P_{m}(k)$, $(dn/dM)_{\Delta}$,
$(dn/dM)_{{\rm g}}$, and $b(k)$ are affected. If one were to simulate
the nonlinear evolution of density perturbations and the structure
formation in $\sim4\,$Mpc patches, he should generate initial conditions
based on this work. As is seen in Figure \ref{fig:Pk_growth}, we
cannot neglect the impact of overdensity even when the simulation
starts at e.g. $z=200$, because at $\sim1\sigma_{\Delta_{{\rm c}}}$
the discrepancy between our prediction and that by TH is already a
few percent at that redshift. 

Both the previous semi-analytical work (\citealt{Tseliakhovich2011};
\citealt{Fialkov2012}; \citealt{McQuinn2012}; \citealt{Bovy2013};
\citealt{Naoz2014}; \citealt{Asaba2016}) and the semi-numerical
work (e.g. \citealt{Fialkov2013}; \citealt{Visbal2012}) should be
re-examined. Attempts to numerically simulate the nonlinear evolution
of small-scale structures have been mostly limited to the physics
inside mean-density regions (\citealt{McQuinn2012}; \citealt{Maio2011};
\citealt{Stacy2011}; \citealt{Greif2011}) or special, isolated regions
(\citealt{Tanaka2014}). These numerical simulations thus need to
be extended to incorporate $\Delta$ which varies in space. In doing
so, a reasonable method would be to use adaptive mesh refinement (AMR)
codes with nested grids, so that one or a few interesting regions
($\sim4\,$Mpc patches with $\Delta=\pm1\sigma_{\Delta}$,
for example) are treated with fine meshes and other regions with coarse
meshes for computational efficiency.

We also showed that cosmology through galaxy surveys should carefully
consider the impact of the mode-mode couping, because the halo bias
(and galaxy bias as well) would be boosted from not only the standard
prediction with $V_{{\rm bc}}=0$ but also the prediction by
TH. Cosmology with the intensity 
mapping may also be affected. The post-reionization intensity mapping
targets the large-angle, diffuse 21-cm background from neutral hydrogen
atoms inside galaxies (\citealt{Chang2008}; \citealt{Abdalla2010};
\citealt{Bandura2014}; \citealt{Xu2015}). Because any galaxies,
small or large, contribute to this cumulative 21-cm background, such
observations will be affected by the streaming velocity through $b(k)$.
The pre-reionization intensity mapping targets the large-angle, diffuse
21-cm background from the intergalactic neutral hydrogen atoms (\citealt{1990MNRAS.247..510S};
\citealt{2004MNRAS.352..142B}; \citealt{Loeb2004}; \citealt{2005ApJ...624L..65B};
\citealt{2006ApJ...653..815M}; ; \citealt{Mao2012}; \citealt{Shapiro2013}).
Even though this is free from the galaxy bias, the streaming velocity
may act as a heating mechanism and boost the power spectrum of the
velocity field (\citealt{McQuinn2012}), and therefore the new findings
of our work should be incorporated.

Application to the study of the cosmic reionization process is of
a prime interest in terms of the high-redshift astrophysics. The complex
nature of the process usually requires numerical simulations, and
they are performed through either efficient semi-numerical methods
(\citealt{2004ApJ...613....1F}; \citealt{McQuinn2007}; \citealt{Mesinger2011};
\citealt{Alvarez2012}) or fully numerical methods (e.g. \citealt{gnedinabel01};
\citealt{2002ApJ...572..695R}; \citealt{2003MNRAS.345..379M}; \citealt{c2ray};
\citealt{Baek2009}; \citealt{Wise2011}). The early phase of cosmic
reionization must have been driven by the first stars, possibly forming
first in minihalos, as these are the first luminous objects in the
universe. A very important factor that modulates the formation of
the first stars inside minihalos is the Lyman-Werner intensity, which
have been properly treated in simulations in a box large enough for
statistical reliability but implementing subgrid physics for the first
star formation inside minihalos (\citealt{Ahn2012}; \citealt{Fialkov2013}).
To predict reionization scenarios to our best knowledge, especially
on its early phase, this work should be properly incorporated because
the star formation inside minihalos is strongly modulated by the streaming
velocity as well.

The results of this paper have rooms for further improvements. This
paper is based on the presumption that the fluctuations at a few Mpc
scale remain linear even at later epochs. However, high density regions
will reach the nonlinear regime earlier than the rest, and then our
formalism will break down in such regions. The halo bias is more strongly
pronounced in the nonlinear patches (e.g. \citealt{Ahn2015}) than
in the linear theory, and thus one should use the actual values of
the overdensity in such circumstances. One could achieve this goal
by adopting the quasi nonlinear calculation (e.g. 2LPT by
\citealt{Crocce2006}), adopting 
the top-hat collapse model as in \citet{Mo1996} and \citet{Ahn2015},
or for the best accuracy running N-body+hydro simulations which resolve
the density fluctuation at $\sim$ Mpc scale. Then, in each patch
of a few comoving Mpc, Equation (\ref{eq:perturbation_k}) can be integrated
with the newly computed values of $\Delta$'s. Wave modes in the range
$k\simeq[1-10]/$Mpc are not accurately treated, because we based
our formalism on the separability of the large-scale modes ($k\lesssim1/$Mpc)
and the small-scale modes ($k\gtrsim10/$Mpc). The code we used will be released for the public use, but it requires technical improvements
such as allowing parallel computation and porting to more generic
computation languages. We will maintain and control its development
through the website \url{http://www.chosun.ac.kr/kjahn}.

\acknowledgments
We thank P. R. Shapiro, F. Schmidt and R. Barkana for helpful
discussions. We also thank the anonymous referee for the clear report which
led to a significant, quantitative improvement of the paper. This
work was supported by a research grant from Chosun University (2016)
and by NRF-2014R1A1A2059811.

\appendix
\section{Normal Modes for the large-scale fluctuations}
When the fluctuation of radiation components are neglected, the
growth of large-scale density and velocity fluctuations are well
approximated by Equations (\ref{eq:background}) and (\ref{eq:background_easy}). 
Their evolution can then
be described by the 4 normal modes described in Section
\ref{sub:perturbation}. 
The growing and decaying modes are the two solutions to the
second-order equation
\begin{equation}
\frac{d^{2}\Delta_{+}}{dt^{2}}+2H\frac{d\Delta_{+}}{dt}-\frac{3}{2}H^{2}\Omega_{m}\Delta_{+}=0,
\label{eq:Deltaplus}
\end{equation}
where $\Delta_{+}=f_{\rm c}\Delta_{\rm c}+f_{\rm b}\Delta_{\rm b}$,
and can be written also as
\begin{equation}
\frac{d^{2}\Delta_{+}}{da^{2}}+\left(\frac{3}{a}+\frac{d\ln H}{da}\right)\frac{d\Delta_{+}}{da}-\frac{3}{2a^{2}}\Omega_{m}\Delta_{+}=0.
\label{eq:Deltaplus_a}
\end{equation}
Similarly, the compensated and streaming modes are the two solutions
to
\begin{equation}
\frac{d^{2}\Delta_{-}}{dt^{2}}+2H\frac{d\Delta_{-}}{dt}=0,
\label{eq:Deltaminus}
\end{equation}
where $\Delta_{-}=\Delta_{\rm c}-\Delta_{\rm b}$,
and can be written also as
\begin{equation}
\frac{d^{2}\Delta_{-}}{da^{2}}+\left(\frac{3}{a}+\frac{d\ln H}{da}\right)\frac{d\Delta_{-}}{da}=0.
\label{eq:Deltaminus_a}
\end{equation}

We now take a convention of writing each mode as the product of
its initial value at $z=1000$ and its growth factor:
$\Delta_{+}^{{\rm g}}(a)=\Delta_{{\rm gro}}D^{{\rm g}}(a)$,
$\Delta_{+}^{{\rm d}}(a)=\Delta_{{\rm dec}}D^{{\rm d}}(a)$,
$\Delta_{-}^{{\rm c}}(a)=\Delta_{{\rm com}}={\rm constant}$, and
$\Delta_{-}^{{\rm s}}(a)=\Delta_{{\rm str}}D^{{\rm s}}(a)$, denoting
the growing, decaying, compensated, and streaming modes,
respectively. These modes comprise $\Delta_{+}$ and $\Delta_{-}$, as 
$\Delta_{+}(a)=\Delta_{{\rm gro}}D^{{\rm g}}(a)+\Delta_{{\rm dec}}D^{{\rm d}}(a)$ and  
$\Delta_{-}(a)=\Delta_{{\rm com}}+\Delta_{{\rm str}}D^{{\rm s}}(a)$,
with the normalization $D^{{\rm g}}=D^{{\rm d}}=D^{{\rm s}}=1$ at
$z=1000$.

\begin{figure}
\includegraphics[width=0.45\textwidth]{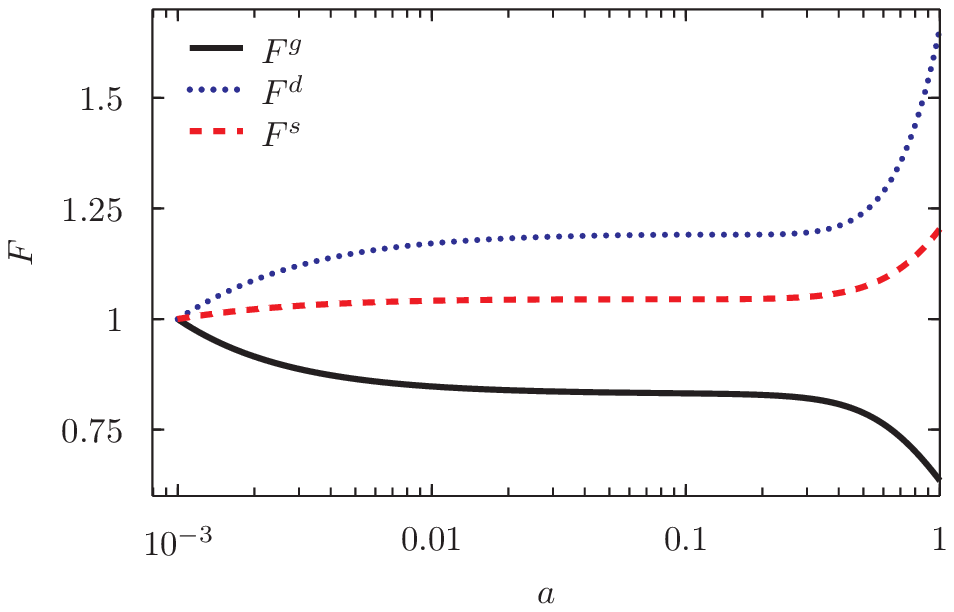}
\includegraphics[width=0.45\textwidth]{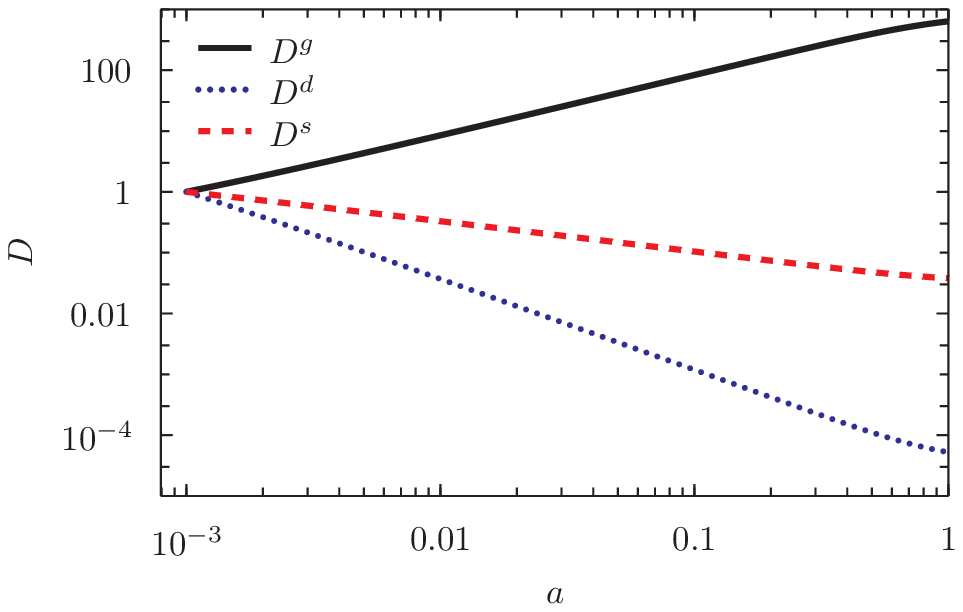}
\caption{Evolution of growth factors, for the growing (black, solid), decaying
(blue, dotted), and streaming (red, dashed) modes. When $F$ (left
  panel) is multiplied to each analytic power-law evolution valid for
  the matter-dominated era, it results in the actual growth factor $D$
(right panel).}
\label{fig:growth}
\end{figure}

The first task in finding these modes is to calculate the growth
factors. Because the Hubble constant
$H(a)$ 
($=H_{0}\sqrt{\Omega_{r,\,0}a^{-4}+\Omega_{m,\,0}a^{-3}+\Omega_{\Lambda,\,0}})$
and $\Omega_{m}(a)$ have non-negligible radiation components during
the period of interest, $1000\gtrsim z \gtrsim 50$, growth factors
should be calculated numerically. We factor out deviations from the
analytical form valid during the matter-dominated
($\Omega_{m}=1$) era, as $D^{{\rm g}}(a)=(a/a_{i})F^{\rm g}(a)$, 
$D^{{\rm d}}(a)=(a/a_{i})^{-3/2}F^{\rm d}(a)$, and 
$D^{{\rm s}}(a)=(a/a_{i})^{-1/2}F^{\rm s}(a)$, and then solve for
the order-of-unity values of
$F^{\rm g}$, $F^{\rm d}$, and $F^{\rm s}$. They are determined by
\begin{equation}
\frac{d^{2}F^{{\rm g}}}{da^{2}}+\left(\frac{5}{a}+\frac{d\ln H}{da}\right)\frac{dF^{{\rm g}}}{da}-\frac{1}{a}\left\{\frac{3}{a}\left(\frac{\Omega_{m}}{2}-1\right)-\frac{d\ln H}{da}\right\} F^{{\rm g}}=0,
\label{eq:Fg}
\end{equation}
\begin{equation}
\frac{d^{2}F^{{\rm d}}}{da^{2}}+\frac{d\ln H}{da}\frac{dF^{{\rm d}}}{da}-\frac{1}{a}\left\{\frac{3}{a}\left(\frac{\Omega_{m}}{2}+\frac{1}{4}\right)+\frac{3}{2}\frac{d\ln H}{da}\right\} F^{{\rm d}}=0,
\label{eq:Fd}
\end{equation}
and
\begin{equation}
\frac{d^{2}F^{{\rm s}}}{da^{2}}+\left(\frac{2}{a}+\frac{d\ln H}{da}\right)\frac{dF^{{\rm s}}}{da}-\frac{1}{a}\left\{ \frac{3}{4a}+\frac{1}{2}\frac{d\ln H}{da}\right\} F^{{\rm s}}=0.
\label{eq:Fs}
\end{equation}
In practice, the numerical integration of Equations
(\ref{eq:Fg})-(\ref{eq:Fs}) is started at $z=10$ (backward for
$z\ge 10$ and forward for $z\le 10$), with the condition
that $dF^{\rm g,\,d,\,s}/da=0$ at $z=10$ because it is a
matter-dominated epoch, and $F^{\rm g,\,d,\,s}=1$ at $z=1000$ because
of our normalization convention. One can instead start the integration
from the radiation dominated epoch (just numerically assuming that
Equations (\ref{eq:Fg})-(\ref{eq:Fs}) are all valid at $z\simeq 10000$ and
using the asymptotes for $dF^{\rm g,\,d,\,s}/da$ and $F^{\rm
  g,\,d,\,s}$ at that time), but we find that Equations (\ref{eq:Fd}) and (\ref{eq:Fs})
become stiff if integrated forward in increasing $a$ at high $z$.
We show the growth factors found this way in Figure \ref{fig:growth}.

\begin{figure}
\center
\includegraphics[width=0.4\textwidth]{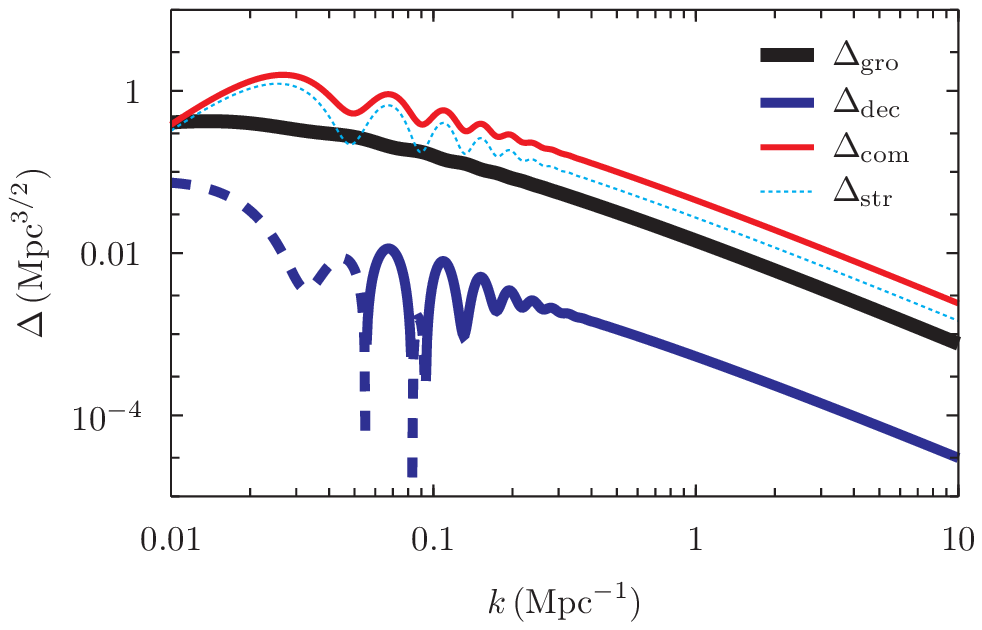}
\caption{Growing, decaying, compensated, and streaming modes at
  $z=1000$, plotted in varying line widths. The negative values are
  plotted by a dashed line for
  the decaying mode and a dotted line for the streaming mode, after
  flipping the sign.}
\label{fig:modes}
\end{figure}

Now we can find the initial values of these modes by using the growth
factors found above on the transfer function outputs from CAMB. We
algebraically relate two redshift outputs, at $z_{1}\equiv z_{i}=1000$ and
$z_{2}\equiv 800$ in practice, to find these modes:
\begin{eqnarray}
\Delta_{{\rm gro}}&=&\frac{\Delta_{+}(a_{2})-D^{{\rm d}}(a_{2})\Delta_{+}(a_{1})}{D^{{\rm g}}(a_{2})-D^{{\rm d}}(a_{2})}, \nonumber \\
\Delta_{{\rm dec}}&=&\Delta_{+}(a_{1})-\Delta_{{\rm gro}}, \nonumber \\
\Delta_{{\rm str}}&=&\frac{\Delta_{-}(a_{2})-\Delta_{-}(a_{1})}{D^{{\rm s}}(a_{2})-1}, \nonumber \\
\Delta_{{\rm com}}&=&\Delta_{-}(a_{1})-\Delta_{{\rm str}},
\label{eq:modes}
\end{eqnarray}
where $\Delta_{\pm}(a_{1,2})$ are those from CAMB.
The modes found this way are shown in Figure \ref{fig:modes}. 

\begin{figure}
\includegraphics[width=0.333\textwidth]{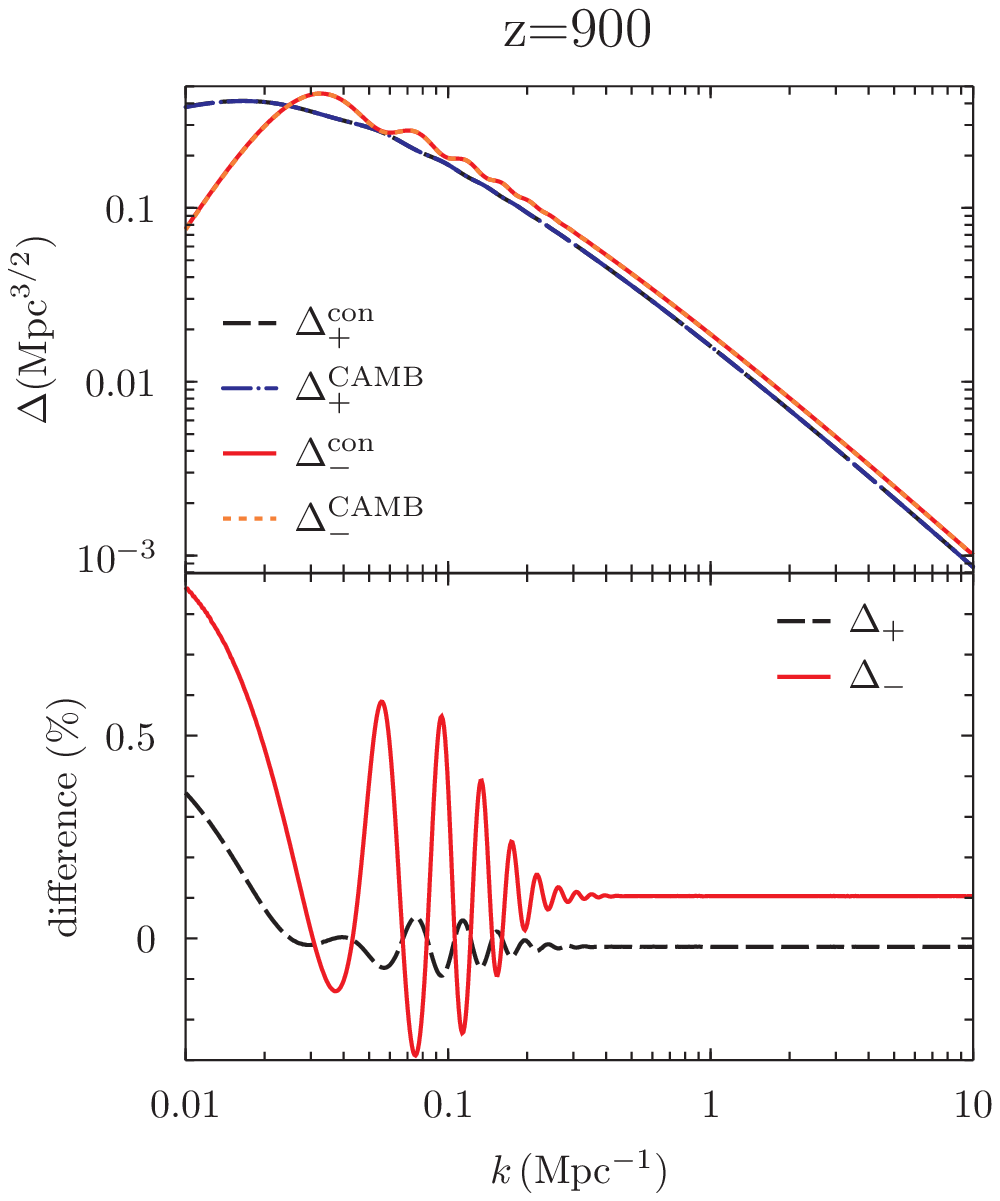}
\includegraphics[width=0.33\textwidth]{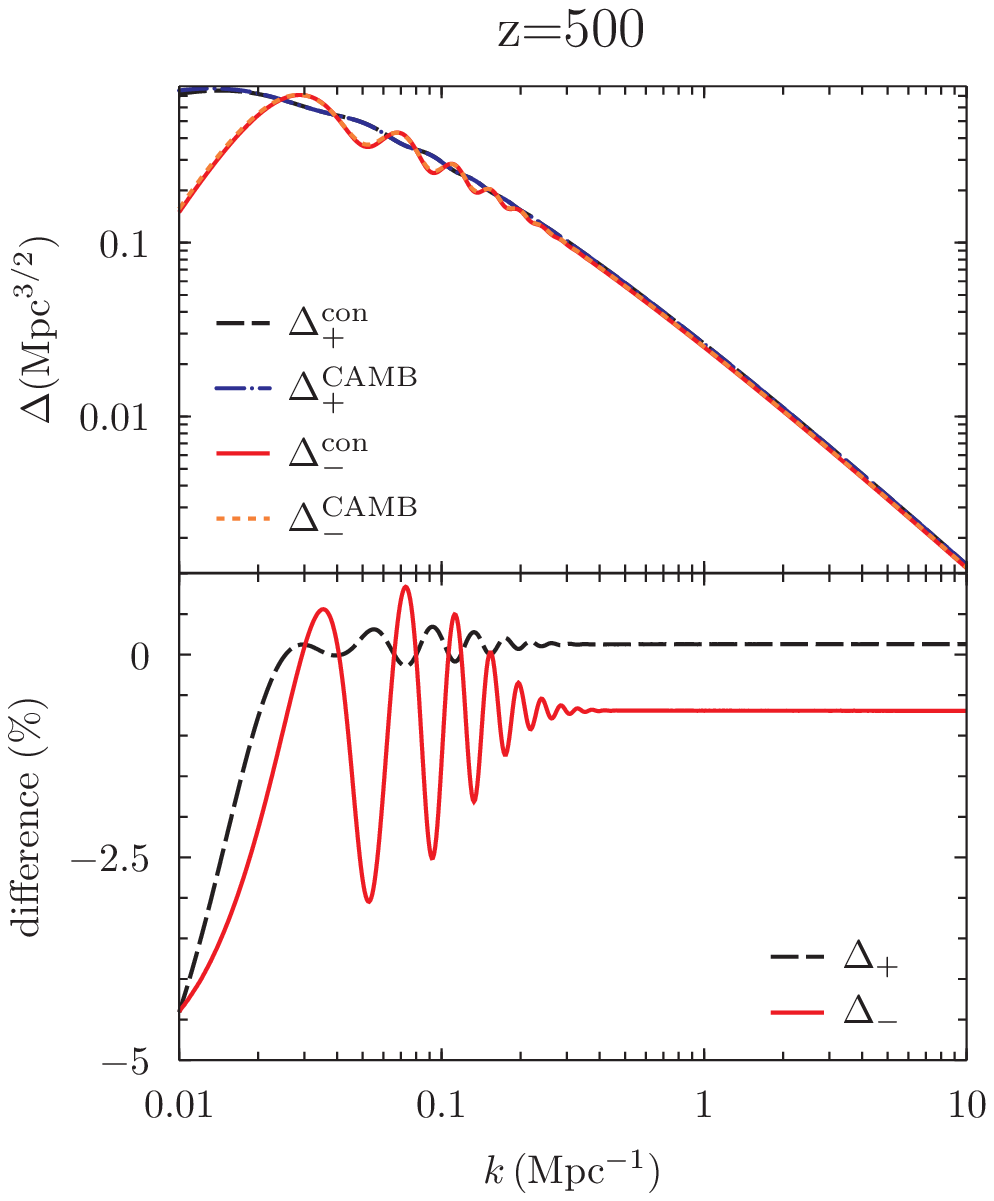}
\includegraphics[width=0.33\textwidth]{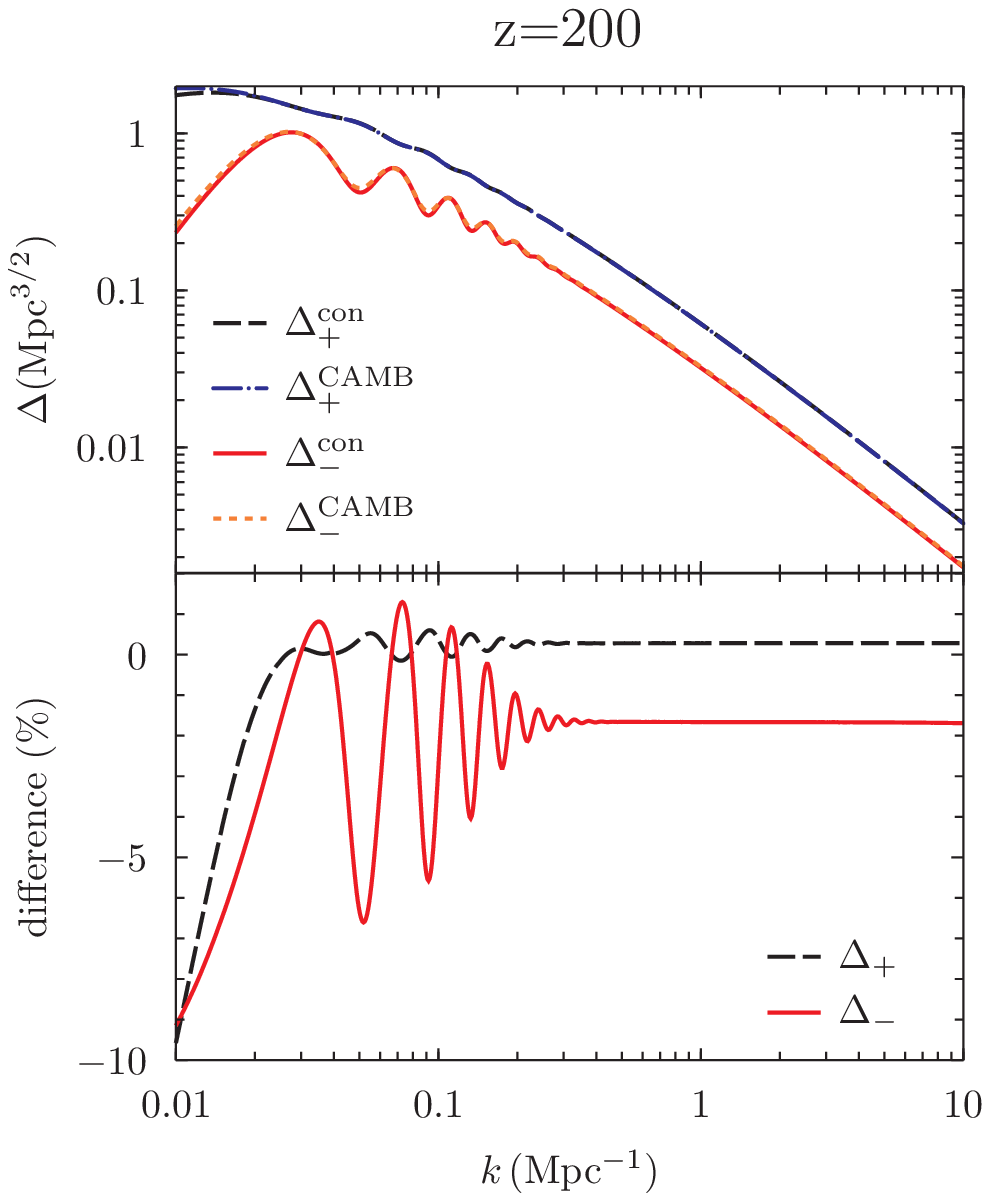}
\caption{Comparison of the density fluctuations constructed by
  the normal modes (with superscript ``con'') and those calculated by
  CAMB (with superscript ``CAMB''), at different redshifts. 
  It is difficult to see the
  difference in the upper panels because the match is good, but
  depending on the value of $k$, some error is
  inherent as seen in the lower panels. The difference is with respect
  to the values calculated by CAMB. We allow this degree of error in
  this paper, but for higher accuracy
  one is advised to use the CAMB outputs for the large-scale fluctuations.}
\label{fig:constructed_vs_CAMB}
\end{figure}

A few things are notable. An
unperturbed Hubble flow requires $\Delta_{\rm gro}/\Delta_{\rm
  dec}=3/2$, while we find that at $z=1000$  $\Delta_{\rm gro}/|\Delta_{\rm
  dec}|\simeq 27$ and is not constant over $k$. Even though one can
choose
different redshifts to extract these modes and they should not change
in principle, the resulting
modes at $z_{i}$ vary depeding on the choice of the redshifts. We
believe that this is partly due to our neglect of the fluctuations of
radiation components, because they can affect the evolution of
density fluctuations. We find that our current choice of $z_{1}$ and
$z_{2}$ is optimal for $k \gtrsim 0.01\,{\rm Mpc}$: using
these modes and evolving them with the 
growth factors, we find a good match between $\{\Delta_{+}$,
$\Delta_{-}\}$ ``constructed'' by using these modes and those calculated by
CAMB, at any redshifts with at most a several percent error. We show their
comparison in Figure \ref{fig:constructed_vs_CAMB}.

\citet{Schmidt2016} follows a similar approach for the mode
extraction but uses CAMB transfer function outputs at $z\simeq
0$. Their focus is on the low-redshift galaxy surveys, and thus the
accuracy is required mostly at and near the present. In
our case, accuracy in $\Delta_{j}$ and $\Theta_{j}$ is required mostly
at a high redshift range, $1000\lesssim z\lesssim 50$, because the
small-scale modes are continuously affected by large-scale
modes from the epoch of recombination and the linear perturbation
analysis on small-scales modes breaks down later when they become nonlinear.  

\section{Fitting formula for the large-scale temperature fluctuation}
For the evolution of $\Delta_{T}$, we integrate Equation
(\ref{eq:Deltatemp}) on each spatial patch. Early on, its value is
strongly affected by the initial fluctuation of the CMB
temperature. Later, it decouples from the CMB fluctuation and is
determined mostly by $\Delta_{\rm b}$. Therefore, one expects a strong
correlation between $\Delta_{T}$ and $\Delta_{\rm b}$ long after
recombination. For example, at $a=0.01$ and $0.1$, they follow the
linear relation: $\Delta_{T}/\Delta_{\rm b}=0.239$ at $a=0.01$, 
and $\Delta_{T}/\Delta_{\rm b}=0.586$ at $a=0.1$. Regardless of the
variance in $\Delta_{T}$, therefore, one can find a fitting formula
for  $\Delta_{T}(a)$ after its decoupling from the CMB temperature.

\begin{figure}
\center
\includegraphics[width=0.4\textwidth]{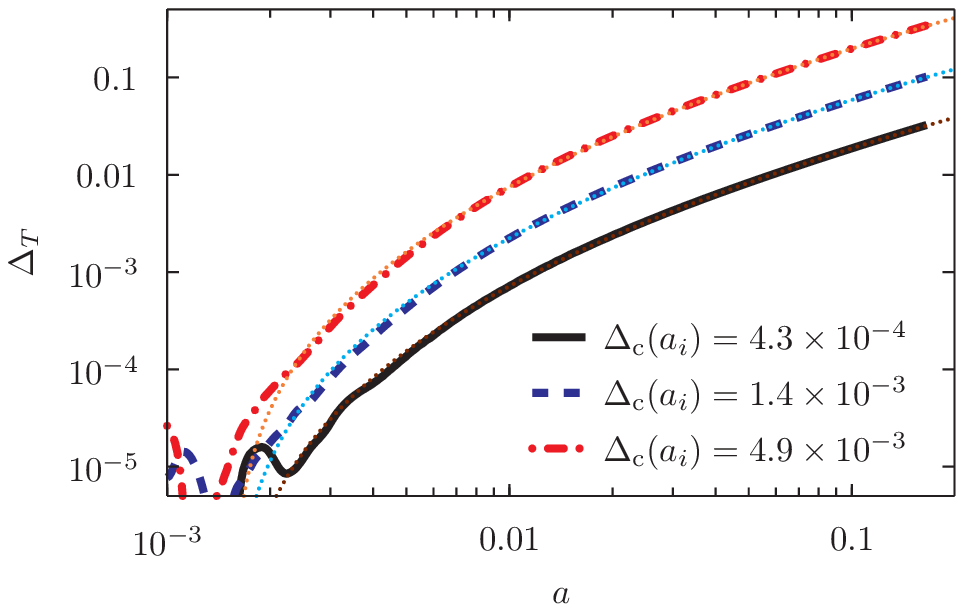}
\caption{Actual evolution of $\Delta_{T}$ of three arbitrarily chosen
  spatial patches (4 comoving Mpc) with varying $\Delta_{\rm
    c}(a_{i})$ (thick; solid, dashed, dot-dashed). Overlaid are the
  corresponding fitting functions given by Equation
  \ref{eq:DeltaT_fitting}, in thin dotted lines. }
\label{fig:DTfit}
\end{figure}

The fitting formula is given by Equation (\ref{eq:DeltaT_fitting}).
We note that 
Equation (\ref{eq:DeltaT_fitting}) is valid only when the patch size
is 4 comoving Mpc. For patches in different size, we believe that a
generic form of a two-parameter fit,
\begin{eqnarray}
\Delta_{T}(a)&=&{\rm sign}(\Delta_{T,\, A})\,{\rm dex}\left[\alpha\left(\log_{10}a+\beta\right)^{\gamma} -Y\right],\nonumber \\
\alpha&=&\frac{\log_{10}\left(\Delta_{T,\, B}/\Delta_{T,\, A}\right)}{(\beta-1)^{\gamma}-(\beta-2)^{\gamma}}, \nonumber \\
Y&=&\frac{ (\beta-2)^{\gamma}\log_{10}|\Delta_{T,\, B}|  -(\beta-1)^{\gamma}\log_{10}|\Delta_{T,\, A}| }{(\beta-1)^{\gamma}-(\beta-2)^{\gamma}},
\label{eq:DeltaT_fit_general}
\end{eqnarray}
will serve as a good fit regardless of the patch size. The linear
relation between $\Delta_{T}$ and $\Delta_{\rm b}$ at $a=0.01$ and
$0.1$ can be found by comparing the two quantities.
In case of the 4 comoving Mpc patch, we find that $\beta=2.8$ and
$\gamma=0.33$ (Equation \ref{eq:DeltaT_fitting}) provides an excellent fit to $\Delta_{T}$ when
$\Delta_{T}\gtrsim 10^{-4}$, with the linear relations
$\Delta_{T}/\Delta_{\rm b}=0.279$ at $a=0.01$ 
and $\Delta_{T}/\Delta_{\rm b}=0.599$ at $a=0.1$ This is demonstrated in Figure
\ref{fig:DTfit}, where the evolution of different $\Delta_{T}$'s
are shown depending on the initial $\Delta_{\rm c}$, together with the
corresponding fits. 

\bibliographystyle{aasjournal}

\end{document}